\documentclass[twocolumn,tighten]{aastex62}

\shorttitle{Hot Gas Gradients in M82}
\shortauthors{LOPEZ ET AL.}

\newcommand{\ltsima}{$\; \buildrel < \over \sim \;$}
\newcommand{\simlt}{\lower.5ex\hbox{\ltsima}}

\def\arcmin{\hbox{$^\prime$}}
\def\arcsec{\hbox{$^{\prime\prime}$}}

\begin{document}

\title{Temperature and Metallicity Gradients in the Hot Gas Outflows of M82}

\correspondingauthor{Laura A. Lopez}
\email{lopez.513@osu.edu}

\author{Laura A. Lopez}
\affil{Department of Astronomy, The Ohio State University, 140 W. 18th Ave., Columbus, OH 43210, USA}
\affil{Center for Cosmology and AstroParticle Physics, The Ohio State University, 191 W. Woodruff Ave., Columbus, OH 43210, USA}
\affil{Niels Bohr Institute, University of Copenhagen, Blegdamsvej 17, 2100 Copenhagen, Denmark}

\author{Smita Mathur} 
\affil{Department of Astronomy, The Ohio State University, 140 W. 18th Ave., Columbus, OH 43210, USA}
\affil{Center for Cosmology and AstroParticle Physics, The Ohio State University, 191 W. Woodruff Ave., Columbus, OH 43210, USA}

\author{Dustin D.~Nguyen}
\affil{Center for Cosmology and AstroParticle Physics, The Ohio State University, 191 W. Woodruff Ave., Columbus, OH 43210, USA}
\affil{Department of Physics, The Ohio State University, 191 W. Woodruff Avenue, Columbus, OH, 43210, USA}

\author{Todd A. Thompson}
\affil{Department of Astronomy, The Ohio State University, 140 W. 18th Ave., Columbus, OH 43210, USA}
\affil{Center for Cosmology and AstroParticle Physics, The Ohio State University, 191 W. Woodruff Ave., Columbus, OH 43210, USA}

\author{Grace M. Olivier}
\affil{Department of Astronomy, The Ohio State University, 140 W. 18th Ave., Columbus, OH 43210, USA}
\affil{Center for Cosmology and AstroParticle Physics, The Ohio State University, 191 W. Woodruff Ave., Columbus, OH 43210, USA}

\begin{abstract}

We utilize deep {\it Chandra} X-ray Observatory imaging and spectra of M82, the prototype of a starbursting galaxy with a multiphase wind, to map the hot plasma properties along the minor axis of the galaxy. We extract spectra from 11 regions up to $\pm$2.5~kpc from the starbursting midplane and model the data as a multi-temperature, optically thin thermal plasma with contributions from a non-thermal (power-law) component and from charge exchange (CX). We examine the gradients in best-fit parameters, including the intrinsic column density, plasma temperature, metal abundances, and number density of the hot gas as a function of distance from the M82 nucleus.  We find that the temperatures and number densities of the warm-hot and hot plasma peak at the starbursting ridge and decreases along the minor axis. The temperature and density profiles are inconsistent with spherical adiabatic expansion of a super-heated wind and suggest mass loading and mixing of the hot phase with colder material. Non-thermal emission is detected in all of the regions considered, and CX comprises $8-25$\% of the total absorption-corrected, broad-band ($0.5-7$~keV) X-ray flux. We show that the abundances of O, Ne, Mg, and Fe are roughly constant across the regions considered, while Si and S peak within 500~pc of the central starburst. These findings support a direct connection between the M82 superwind and the warm-hot, metal-rich circumgalactic medium (CGM). 

\end{abstract}


\section{Introduction} 

Galaxy-scale outflows driven by star formation are ubiquitous \citep{heckman90,veilleux05,rubin14}. These galactic winds chemically enrich the circumgalactic (CGM) and intergalactic medium (IGM; \citealt{bor13,werk16}) as well as regulate the growth and metal-enrichment of galactic disks \citep{oppenheimer08,peeples11}. The prevailing picture is that outflows are driven by hot gas shock-heated by stellar winds and supernovae (SNe) that entrain the dust, cold, and warm gases within the flow (e.g., \citealt{chevalier85}). Additional mechanisms that have been proposed to accelerate cool clouds in galactic winds include radiation pressure of starlight on dust grains (e.g., \citealt{murray11,thompson15}) and cosmic rays (e.g., \citealt{ipavich75,everett08,Socrates2008,booth13}).

M82 is the prototype of a starbursting galaxy \citep{rieke80} with a strong, multiphase wind \citep{leroy15}. It is located only 3.6~Mpc away \citep{freedman94,gerke11} and is nearly edge on (with a disk inclination of 80$^{\circ}$: \citealt{mckeith95}), making it well-suited to observe the biconical outflows along its minor axis. The outflows of M82 have been studied across the electromagnetic spectrum, tracing the $\lesssim$100~K atomic H~{\sc i} and molecular gas \citep{walter02,salak13,beirao15,martini18}, the $\sim$10$^{4}$~K warm-ionized gas in H$\alpha$ \citep{mckeith95,westmoquette09}, the $\sim$10$^{7}$~hot gas in X-rays \citep{watson84,bregman95,strickland97}, and the entrained dust in the UV, IR, and sub-mm \citep{hoopes05,leeuw09,kaneda10,roussel10}. Detailed comparison of the multi-phase wind has revealed that the H~{\sc i} and CO confine the hot outflow \citep{leroy15}, and the H$\alpha$ is well correlated with the diffuse X-rays, with the latter tending to be upstream or interior to the H$\alpha$ \citep{shopbell98,lehnert99,heckman17}. 

Many past X-ray studies of M82 have aimed to measure the properties of the hot gas that entrains the colder components. Several works \citep{ptak97,tsuru97,umeda02} analyzed {\it ASCA} data and found the best-fit model is comprised of two thermal plasmas with enhanced $\alpha$ elements relative to Fe. \cite{cappi99} found similar results using {\it BeppoSAX} data. \cite{read02} analyzed {\it XMM-Newton} Reflection Grating Spectrometers (RGS) data and found near-solar abundances of Fe and O and supersolar abundances of Mg and S. \cite{origlia04} analyzed {\it Chandra} observations as well as {\it XMM-Newton} RGS data, and they found that the hot gas in the nuclear region of M82 was enhanced in $\alpha$ elements relative to Fe. 

\cite{strickland07} analyzed {\it Chandra} and {\it XMM-Newton} observations, focusing on hard ($>$3~keV) X-rays, and showed that 20--30\% of the total hard X-ray luminosity $L_{X}^{2-8~{\rm keV}}$ $\sim1.5\times10^{40}$~erg~s$^{-1}$ arises from diffuse gas and that the continuum likely has a non-thermal component. \cite{strickland09} considered the same X-ray data to constrain the efficiency of SN thermalization and the mass loading, restricting their analysis to the central 500~pc of M82. Using their X-ray spectral results as inputs for hydrodynamical simulations, they found evidence of high thermalization efficiency and mild mass loading.

While extensive X-ray studies have been conducted, most work focuses on the nuclear region of M82 or on the integrated spectra from the disk and halo together. Two exceptions are the works of \cite{ranalli08} and \cite{konami11}. \cite{ranalli08} performed a detailed analysis of how the hot plasma properties vary along M82's minor axis using a 73-ks {\it XMM-Newton} observation. They showed a two-temperature plasma was necessary, with the hot component temperature (of $\sim$7~keV $\approx8\times10^{7}$~K) staying relatively constant with distance from the disk, whereas the warm-hot component decreased from 0.53~keV ($\approx6\times10^{6}$~K) to 0.35~keV ($\approx4\times10^{6}$~K) from the midplane to $\pm$2~kpc along the minor axis. Additionally, \cite{ranalli08} found that many elements (O, Ne, Mg, and Fe) were more abundant in the outflows than in the disk, with this result less pronounced for Si and reversed for S.

\cite{konami11} also considered the temperature and abundance gradients in the M82 outflows using 101 ks of {\it Suzaku} observations. They extracted spectra from the M82 disk and three regions north of the disk, and they modeled the data as a multi-temperature plasma. Their best-fit models included a softer component (of temperature $\sim$0.25~keV) than \cite{ranalli08}, and they found super-solar abundance ratios of O/Fe, Ne/Fe, and Mg/Fe that did not vary spatially with distance from the disk. 

In this paper, we perform a similar analysis as \cite{ranalli08} and \cite{konami11} on deep {\it Chandra} observations of M82. We construct high-resolution images of the diffuse X-ray emission, and we explore how the hot plasma properties change along M82's minor axis out to $\pm$2.5~kpc from the starbursting ridge \citep{lester90}. These {\it Chandra} data have been analyzed previously to study an ultra-luminous X-ray source (ULX; \citealt{brightman16}), to constrain the progenitor of SN~2014J \citep{nielsen14}, and to consider the contribution of charge exchange to the soft X-ray flux of M82 \citep{zhang14}. The {\it Chandra} data considered here are $\sim$10$\times$ deeper than those analyzed by \cite{origlia04} and \cite{strickland09}, and the superb spatial resolution of {\it Chandra} (with an on-axis point spread function [PSF] of 0.492\arcsec) facilitates reliable removal of point sources to study the faint diffuse X-rays of the M82 biconical outflows. 

\begin{deluxetable}{lrc}
\tablecolumns{3}
\tablewidth{0pt} \tablecaption{{\it Chandra} Observations \label{table:data}} 
\tablehead{\colhead{ObsID} & \colhead{Exposure} & \colhead{UT Start Date}}  
\startdata
361 & 33~ks & 1999-09-20 \\
1302 & 16~ks & 1999-09-20 \\
2933 & 18~ks & 2002-06-18 \\
10542 & 119~ks & 2009-06-24 \\
10543 & 118~ks & 2009-07-01 \\
10544 & 74~ks & 2009-07-07 \\
10545 & 45~ks & 2009-07-07 \\
10925 & 95~ks & 2010-07-28 \\
11800 & 17~ks & 2010-07-20
\enddata
\end{deluxetable}

The paper is structured as follows. In Section~\ref{sec:data}, we outline the {\it Chandra} observations and the analysis to produce images and spectra of the diffuse emission. In Section~\ref{sec:results}, we present the results, focusing on the gradients in the temperature and metal abundances as a function of distance from the starburst. We find broad agreement with previous works on the temperature and density of the central region, but our abundance profiles in the outflows differ qualitatively and quantitatively from those of \cite{ranalli08}. In Section~\ref{sec:discussion}, we discuss the implications regarding the mass loading and chemical enrichment of the outflows. In Section~\ref{sec:conclusions}, we summarize our conclusions and paths for future work. We adopt a distance of 3.6~Mpc to M82 throughout this paper; in this case, 1\arcmin\ $\approx$ 1~kpc.

\section{Observations and Data Analysis} \label{sec:data}

\begin{figure}
\begin{center}
\includegraphics[width=\columnwidth]{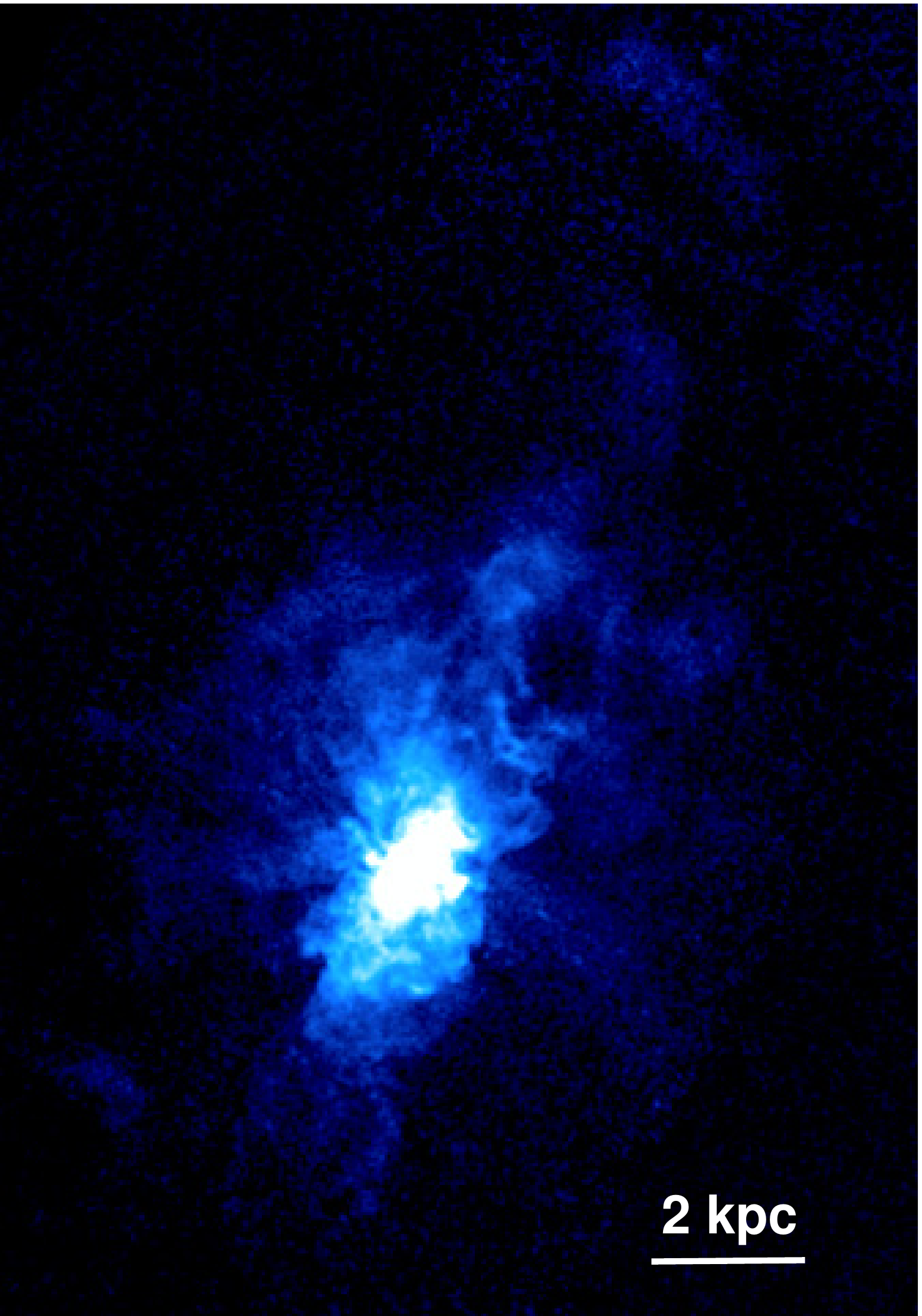}
\end{center}
\vspace{-3mm}
\caption{Exposure-corrected broad-band ($0.5-7.0$~keV) X-ray image of M82 of the diffuse gas with point sources removed. The outflow extends south $\gtrsim$6~kpc from the disk and up to the `Cap' $\sim$12~kpc north of the disk (the diffuse substructure at the top right of the image). North is up, and East is left.}
\label{fig:image}
\end{figure}

We analyze nine {\it Chandra} ACIS-S observations from 1999 (PI: G. Garmire), 2002 (PI: D. Strickland), and 2009--2010 (PI: D. Strickland) listed in Table~\ref{table:data} that total 534~ks. We reduced the data using the {\it Chandra} Interactive Analysis of Observations ({\sc ciao}) Version 4.7 and produced exposure-corrected images using the {\sc ciao} command {\it merge\_obs}. Read-out streaks from ULX M82 X-1 were removed using the {\sc ciao} command {\it acisreadcorr}\footnote{https://cxc.harvard.edu/ciao/threads/acisreadcorr}. Point sources were identified with {\it wavdetect} on the merged, broad-band ($0.5-7.0$~keV) image, incorporating the PSF map of each observation from {\it mkpsfmap}. These point sources were subsequently removed with {\it dmfilth} to construct the image of M82's diffuse gas (Figure~\ref{fig:image}). 

\begin{figure}
\begin{center}
\includegraphics[width=\columnwidth]{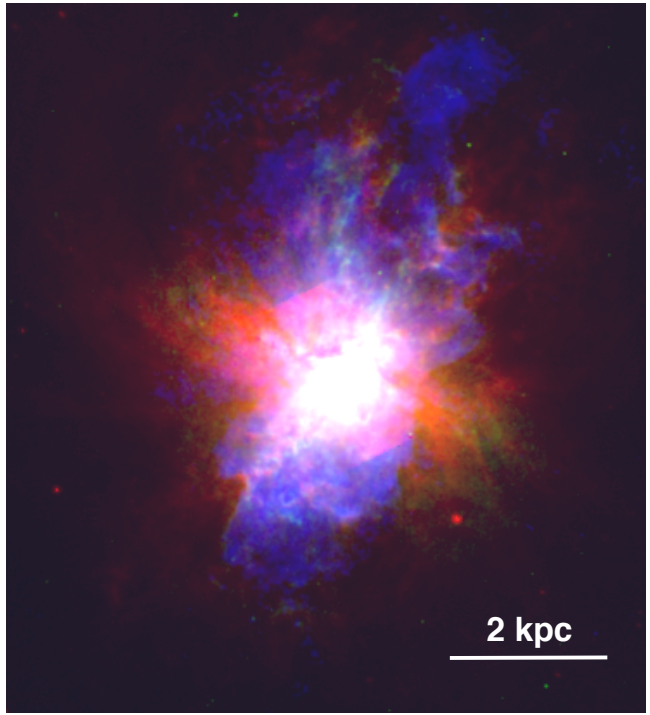}
\end{center}
\vspace{-3mm}
\caption{Three-color image of M82, with 8-$\mu$m in red \citep{engel06}, H$\alpha$ in green \citep{kenn08}, and $0.5-7$.0~keV X-rays in blue. White colors mean the three wavebands are all luminous in that location, and cyan colors represent areas with bright X-rays and H$\alpha$.} North is up, and East is left.
\label{fig:threecolor}
\end{figure}

As shown in Figure~\ref{fig:image}, the diffuse gas extends $\gtrsim$6~kpc south of the disk and north up to the `Cap', a region of diffuse X-rays $\sim$12~kpc north of M82 \citep{lehnert99,tsuru07}. Figure~\ref{fig:threecolor} is a three-color image of M82, with 8-$\mu$m in red \citep{engel06}, H$\alpha$ in green \citep{kenn08}, and the broad-band X-rays in blue. The H$\alpha$ is well correlated with the diffuse X-rays, with the latter being upstream or interior to the H$\alpha$, as noted by previous works \citep{shopbell98,lehnert99,heckman17}. 

\begin{figure*}
\begin{center}
\includegraphics[width=0.7\textwidth]{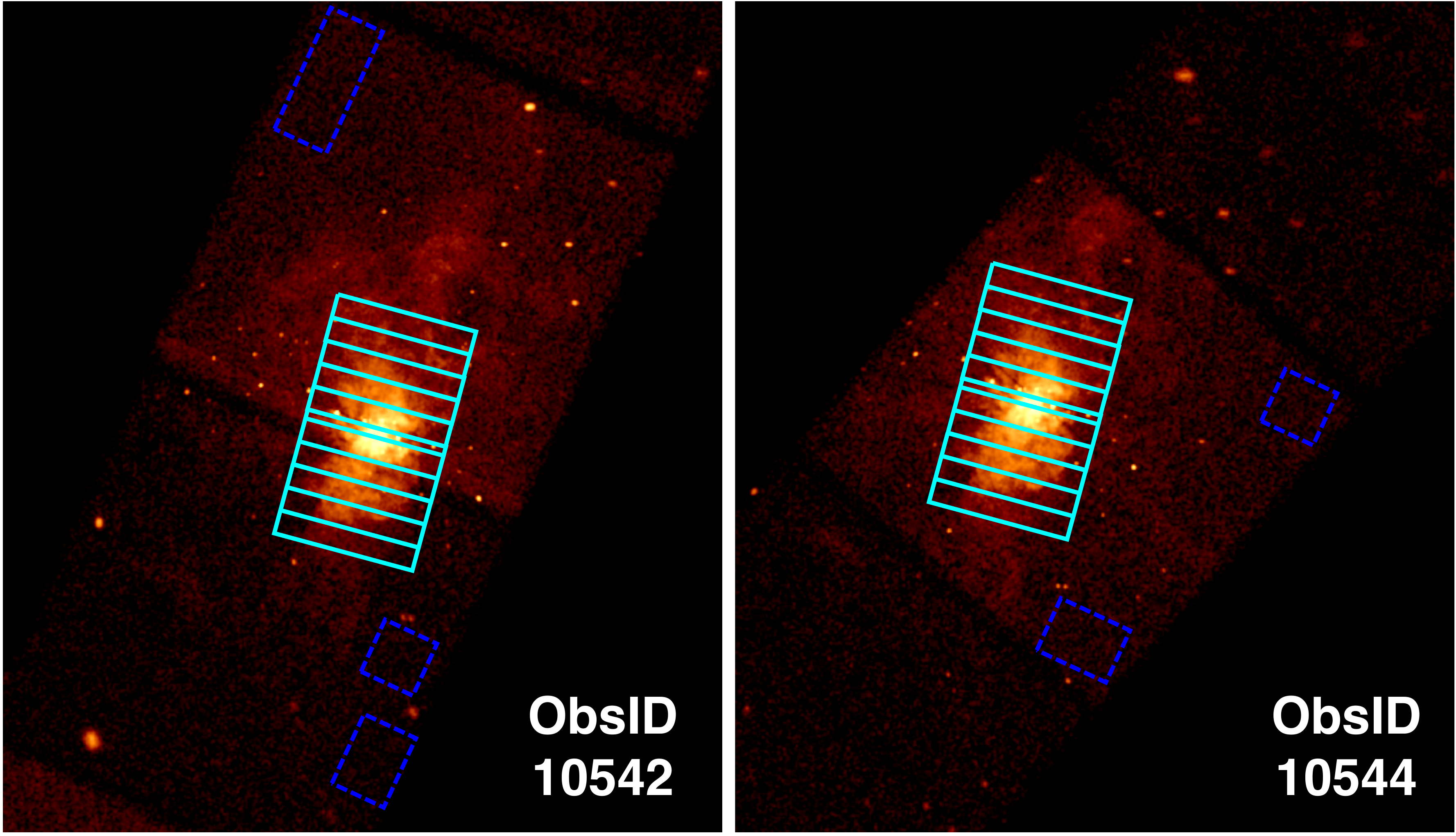}
\end{center}
\vspace{-3mm}
\caption{Broad-band ($0.5-7.0$~keV) images from ObsIDs 10542 (left) and 10544 (right) with source regions (in cyan) and background regions (in blue) overplotted that were used in the spectral analysis. For ObsIDs 10542 and 10543, seven source regions were on the ACIS-S3 chip, while four source regions were on the ACIS-S2 chip. Thus, we adopted background regions that were on the same chips as the source regions being considered.}
\label{fig:background}
\end{figure*}

\begin{figure*}[t]
\begin{center}
\includegraphics[width=\textwidth]{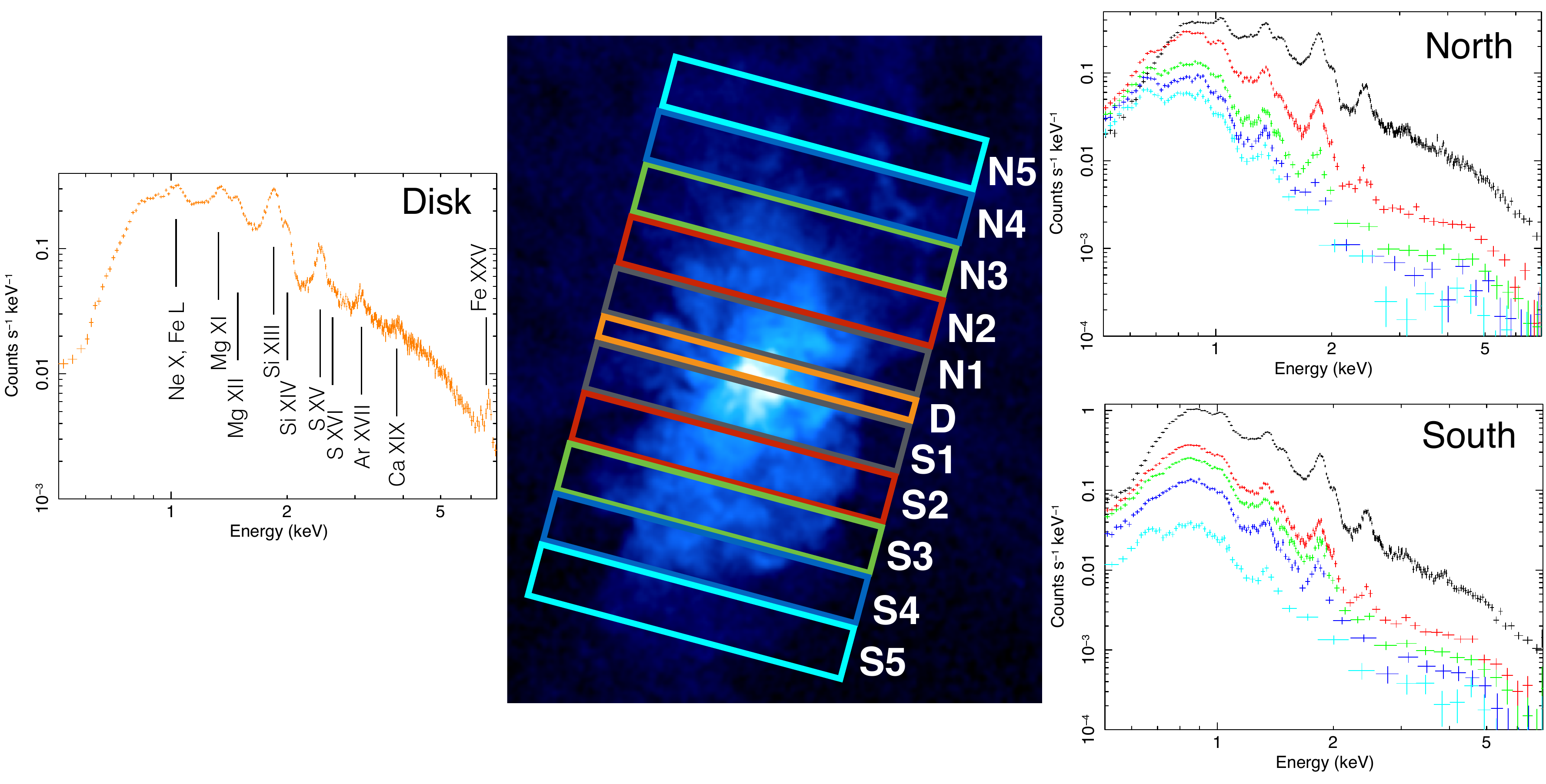}
\end{center}
\vspace{-5mm}
\caption{Spectra were extracted from the 11~regions in the middle panel. Region D (in orange; aligned with the starbursting ridge) is 0.2\arcmin$\times$3\arcmin\ in size, and the outflow regions are 0.5\arcmin$\times$3\arcmin\ in size (recall at the distance of M82, 1\arcmin\ $\approx$ 1 kpc). {\bf Left}: Combined spectrum from region D. Prominent emission lines are labeled; O {\sc viii} (at $\approx$0.65~keV) is not detected because of the high intrinsic column density of $N_{\rm H}^{\rm M82}=(7.8\pm0.2)\times10^{21}$~cm$^{-2}$ there. {\bf Right}: Combined spectra from the North (top) and South (bottom) outflows. Spectra are plotted in the same color as the region box (middle panel) denoting where the data were extracted.}
\label{fig:spectra}
\end{figure*}

To assess the conditions of the diffuse hot gas, we extracted spectra using the {\sc ciao} command {\it specextract} from a 0.2\arcmin$\times$3.0\arcmin\ region (with an area of 2160 arcsec$^{2}$) aligned with the M82 major axis (centered on the position of the M82 nucleus and oriented along the starburst ridge: \citealt{lester90}) as well as ten 0.5\arcmin$\times$3.0\arcmin\ regions  (five north and five south of the disk with areas of 5400 arcsec$^{2}$ each; as shown in Figure~\ref{fig:background} and labeled in Figure~\ref{fig:spectra}). In our spectral analysis, we considered only the six 2009--2010 observations (totaling 467~ks) to limit systematic differences arising from different chip configurations and temporal variations in the years between observations. We excluded all point sources identified by {\it wavdetect} within our regions. 

For each observation, background spectra were extracted from 1--3 selected regions (see Figure~\ref{fig:background} for examples) with a total area of 2.8\arcmin$\times$1.2\arcmin\ ($=$12096 arcsec$^{2}$). For observations with ObsIDs 10542 and 10543, seven of the source regions (S1, D, and N1--N5) are located on the S3 chip, while four of the source regions (S2--S5) are on the S2 chip. Thus, we were careful to adopt background regions on the same chips as their source regions. We note that to avoid the diffuse emission of M82, the background regions tend to be farther off-axis than the source regions: the former are up to 6\arcmin\ off-axis, while the latter are up to 4\arcmin\ off-axis. However, the effective area 6\arcmin\ off-axis at 1.5~keV is $\approx$93\% of the on-axis effective area\footnote{See Figure 4.5 of the {\it Chandra} Proposers' Observatory Guide: https://cxc.harvard.edu/proposer/POG/html/index.html}. Since the emission in the source regions is $\sim3-30\times$ brighter than the background, the diminished off-axis count rate of the background regions has negligible effect on the best-fit spectral parameters.

Background spectra were subtracted from source spectra and were modeled using XSPEC Version 12.10.1 \citep{arnaud96}. We fit the data from each observation jointly by including a multiplicative factor (with the XSPEC component \textsc{const}) that was allowed to vary while all other model parameters were required to be the same between observations. This factor accounts for slight changes in the flux/emission measure between the observations and varied by only $\lesssim$2\% in the best spectral fits. We included two absorption components (with the XSPEC components \textsc{phabs} and \textsc{vphabs}, respectively): one to account for the Galactic absorption of $N_{\rm H} = 4.0\times10^{20}$~cm$^{-2}$ \citep{dickey90} in the direction toward M82 and another to represent M82's intrinsic absorption $N_{\rm H}^{\rm M82}$ which was allowed to vary and had abundances of $Z_{\sun}$ \citep{origlia04}. We adopted cross-sections from \cite{verner96} and solar abundances from \cite{asplund09}.

We began by fitting the spectra with a single, absorbed optically-thin thermal plasma component in collisional ionization equilibrium (CIE) with variable abundances (\textsc{vapec}; \citealt{foster12}) and temperature $T_{1}$. However, we found large residuals associated with line emission (particularly Mg~{\sc xii} and Si~{\sc xiv}) and at hard ($>$2~keV) X-ray energies that improved with the addition of another thermal and a non-thermal component. Specifically, F-tests showed that a power-law ({\sc powerlaw}) component statistically significantly improved the fits in all 11 regions. F-tests also demonstrated that a second, hot thermal plasma component with temperature $T_{2}$ was necessary in 8 of the 11 regions. In all regions, the photon index $\Gamma$ was frozen to $\Gamma=1.5$ as it was not well constrained but was necessary to model accurately the spectra above 2~keV. When allowed to vary, best-fit $\Gamma$ values of $\Gamma = 1-2$ were obtained in the 11 regions, leading us to adopt $\Gamma = 1.5$. We note that in their fit of the central M82 region, \cite{ranalli08} did not exclude point sources and found a best-fit $\Gamma=1.60^{+0.04}_{-0.03}$.

In addition to the model components described above, we included the AtomDB charge-exchange (CX) model component \textsc{vacx}\footnote{http://www.atomdb.org/CX/} to account for line emission produced when ions capture electrons from neutral material \citep{smith12}. A CX component is necessary because previous studies of {\it XMM-Newton} RGS data have demonstrated that $\sim$25\% of the $0.4-2$~keV flux \citep{zhang14} and 50\% of the flux from O {\sc vii}, Ne {\sc ix}, and Mg {\sc xi} originates from CX \citep{liu11}. We tied the individual abundances together in the CX and two thermal components, and we let the abundances of metals with detected emission lines (O, Ne, Mg, Si, S, and Fe) vary. The abundances of metals not detected in the $0.5-7$~keV band were set to 1~$Z_{\sun}$, consistent with the M82 disk metallicity \citep{origlia04}. Given the elevated $N_{\rm H}^{\rm M82}$ in the three central regions S1, D, and N1, the detection of O was limited there, and we froze the O abundance to solar metallicity in those locations.

Putting all of the model components together, the complete XSPEC model for regions S1--S4, and N1--N3 was: \textsc{const*phabs*vphabs*(vapec+vapec+vacx}  \textsc{+powerlaw)}. The models for the southern-most and two northern-most outflow regions (S5, N4--N5) did not include the second \textsc{vapec} component as they did not statistically improve the fits. As detailed below, a third \textsc{vapec} component was added to region~D to account for the detected Fe~{\sc xxv} line. 

\begin{deluxetable*}{lccccccccccr}
\tablecolumns{12}
\tablewidth{0pt} \tablecaption{Spectral Fit Results\tablenotemark{a} \label{table:fitresults}} 
\tablehead{\colhead{Reg.} & \colhead{$r$\tablenotemark{b}} & \colhead{$N_{\rm H}^{\rm M82}$} & \colhead{$kT_{1}$} & \colhead{$kT_{2}$} & \colhead{O/O$_{\sun}$\tablenotemark{c}} & \colhead{Ne/Ne$_{\sun}$} & \colhead{Mg/Mg$_{\sun}$} & \colhead{Si/Si$_{\sun}$\tablenotemark{d}} & \colhead{S/S$_{\sun}$\tablenotemark{d}} & \colhead{Fe/Fe$_{\sun}$} & \colhead{$\chi^{2}$/d.o.f.} \\
\colhead{} & \colhead{(kpc)} & \colhead{($\times10^{21}$~cm$^{-2}$)} & \colhead{(keV)} & \colhead{(keV)} & \colhead{} & \colhead{} & \colhead{} & \colhead{} & \colhead{} & \colhead{}
}  
\startdata
N5 & 2.5 & $<$1.0 & 0.37$\pm$0.02 & -- & 0.7$^{+0.3}_{-0.2}$  & 1.2$^{+0.5}_{-0.3}$ & 0.7$^{+0.3}_{-0.2}$ & 1.0 & 1.0 & 0.23$^{+0.08}_{-0.05}$ & 752/775 \\
N4 & 1.9 & 0.6$\pm$0.4 & 0.37$\pm$0.01 & -- & 0.9$^{+0.3}_{-0.2}$ & 1.7$^{+0.6}_{-0.4}$ & 0.9$^{+0.3}_{-0.2}$ & 1.0 & 1.0 & 0.28$^{+0.08}_{-0.06}$ & 880/861 \\
N3 & 1.4 & $<$0.6 & 0.38$^{+0.01}_{-0.02}$ & 0.86$^{+0.06}_{-0.09}$ & 1.5$^{+0.6}_{-0.4}$ & 2.6$^{+1.1}_{-0.7}$ & 1.7$^{+0.7}_{-0.5}$ & 1.2$^{+0.5}_{-0.3}$  & 1.0 & 0.55$^{+0.21}_{-0.13}$ & 942/952 \\
N2 & 0.9 & 2.2$\pm$0.3 & 0.42$\pm$0.01 & 0.77$^{+0.06}_{-0.05}$ & 0.8$^{+0.2}_{-0.1}$ & 1.6$^{+0.3}_{-0.3}$ & 1.0$\pm$0.2 & 1.1$^{+0.2}_{-0.1}$ & 0.7$\pm$0.3 & 0.38$^{+0.06}_{-0.05}$ & 1278/1191 \\
N1 & 0.4 & 5.9$\pm$0.2 & 0.65$\pm$0.01 & 1.72$^{+0.13}_{-0.05}$ & 1.0 & 1.6$\pm$0.1 & 1.3$\pm$0.1 & 1.8$\pm$0.1 & 2.3$\pm$0.2 & 0.35$\pm$0.03 & 2282/1907 \\
D\tablenotemark{e} & 0 & 7.7$\pm$0.3 & 0.72$\pm$0.02 & 1.46$^{+0.13}_{-0.25}$ & 1.0 & 1.5$\pm$0.2 & 1.5$\pm$0.2 & 2.4$^{+0.3}_{-0.2}$ & 3.6$^{+0.4}_{-0.3}$ & 0.52$^{+0.06}_{-0.05}$ & 2524/2111 \\
S1 & $-$0.4 & 4.2$\pm$0.1 & 0.61$\pm$0.02 & 0.97$^{+0.06}_{-0.04}$ & 1.0 & 1.3$\pm$0.1 & 1.0$\pm$0.1 & 0.9$\pm$0.1 & 1.0$\pm$0.1 & 0.42$\pm$0.02 & 2797/1724 \\
S2 & $-$0.9 & 0.9$\pm$0.3 & 0.48$\pm$0.03 & 0.91$^{+0.03}_{-0.04}$ & 0.5$\pm$0.1 & 1.2$\pm$0.2 & 1.0$^{+0.2}_{-0.1}$ & 0.9$\pm$0.1 & 0.4$\pm$0.2 & 0.51$^{+0.08}_{-0.07}$ & 1157/1093 \\
S3 & $-$1.4 & 0.6$\pm$0.4 & 0.45$\pm$0.03 & 0.87$^{+0.05}_{-0.08}$ & 0.4$\pm$0.1 & 1.1$\pm$0.2 & 0.8$^{+0.2}_{-0.1}$ & 0.7$\pm$0.1 & $<$0.5 & 0.39$^{+0.09}_{-0.05}$ & 1060/969 \\
S4 & $-$1.9 & $<$0.4 & 0.46$\pm$0.03 & 0.86$^{+0.07}_{-0.13}$ & 0.5$\pm$0.1 & 1.3$^{+0.3}_{-0.2}$ & 0.9$\pm$0.2 & 0.7$\pm$0.2 & $<$0.5 & 0.40$^{+0.12}_{-0.07}$ & 917/863 \\
S5 & $-$2.5 & 1.2$^{+0.7}_{-0.6}$ & 0.38$^{+0.02}_{-0.03}$ & -- & 0.8$^{+0.4}_{-0.3}$ & 1.2$^{+0.6}_{-0.4}$ & 0.5$^{+0.3}_{-0.2}$ & 1.0 & 1.0 & 0.25$^{+0.11}_{-0.07}$ & 684/691 \\
\enddata
\tablenotetext{a}{Results reported here are the best-fit parameters of the {\sc vphabs} and {\sc vapec} components of the XSPEC model described in Section~\ref{sec:data}.}
\tablenotetext{b}{$r$ is the distance from the starburst ridge, along the minor axis. Positive values are North of the midplane; negative values are South.}
\tablenotetext{c}{High $N_{\rm H}^{\rm M82}$ in regions D, N1, and S1 precludes detection of the oxygen lines. Thus, the oxygen abundance for these regions is frozen to solar values.}
\tablenotetext{d}{In regions where Si and S lines are not statistically significantly detected, the abundances are frozen to solar values.}
\tablenotetext{e}{A third {\sc vapec} component not listed here was added to the region D model to represent the very hot component. The best-fit temperature of that component was $kT_{3} = 6.89^{+1.83}_{-0.86}$~keV.}
\vspace{-5mm}
\end{deluxetable*}

\section{Results} \label{sec:results}

Figure~\ref{fig:spectra} shows the extracted spectra from the 11~regions, with data from all six observations combined using the {\sc ciao} command {\it combine\_spectra}\footnote{Note that we show the combined spectra as a qualitative demonstration of the detected features. However, in our spectral analysis, we opted to fit the six observations simultaneously (i.e., without combining the data) due to uncertainties introduced by the combining process. See the caveats outlined at https://cxc.harvard.edu/ciao/ahelp/combine\_spectra.html.}. In the spectrum from the disk (region~D), prominent emission lines are evident from Ne, Mg, Si, S, Ar, Ca, and Fe. Among these features, a Fe {\sc xxv} line is apparent at $\approx$6.67$\pm$0.02~keV line (based on a Gaussian fit to that feature) that is not detected in the outflow regions, consistent with \cite{strickland07} who found that it extends $<$100~pc along the M82 minor axis. To account for the Fe {\sc xxv} line, we added a third {\sc vapec} component to the region D model that yielded a best-fit very hot temperature of $kT_{3} = 6.89^{+1.83}_{-0.86}$~keV.

\begin{figure*}[t]
\begin{center}
\includegraphics[width=\textwidth]{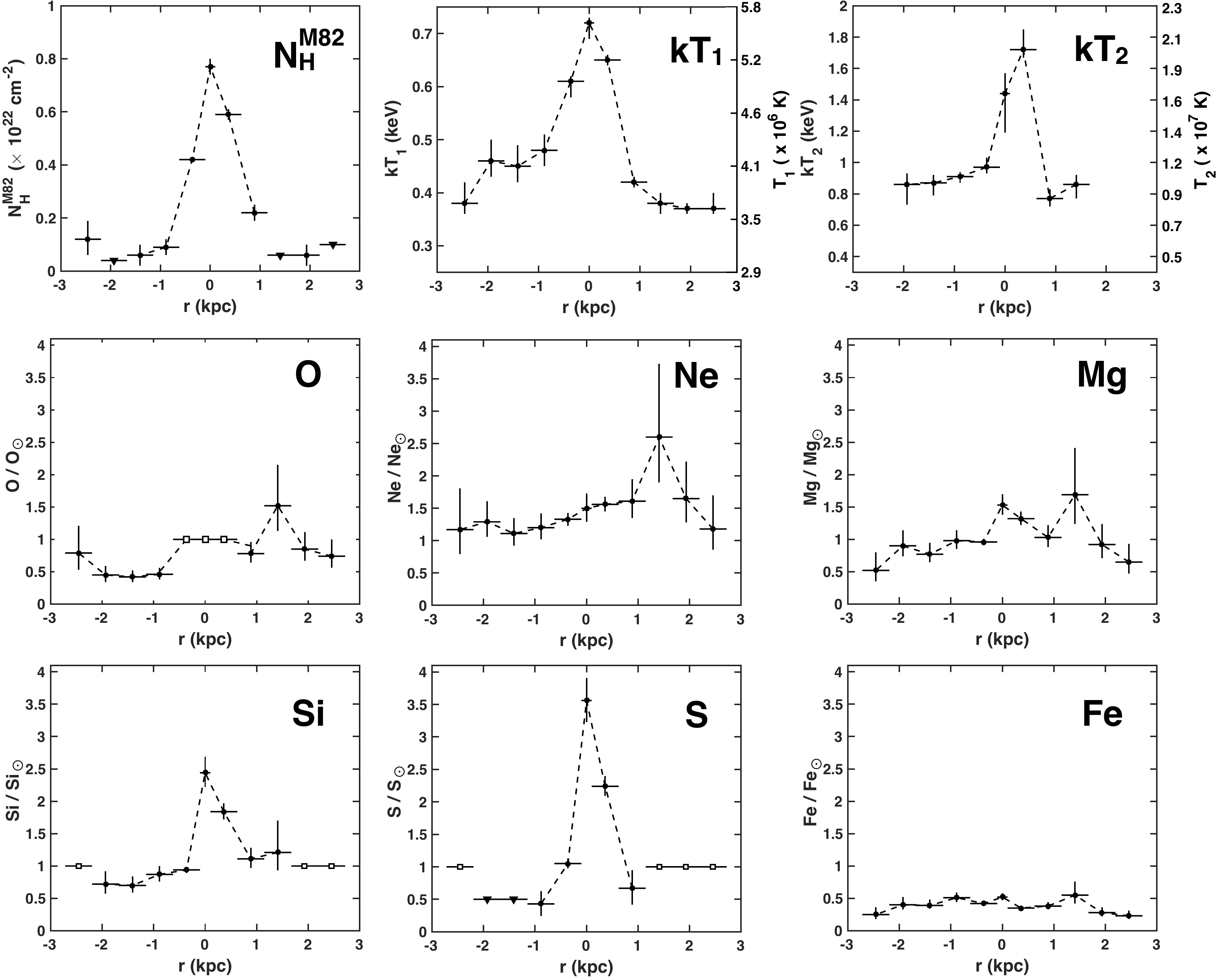}
\end{center}
\vspace{-3mm}
\caption{Best-fit parameters for the 11 regions plotted as a function of $r$, the distance from the M82 major axis. A distance of zero is defined as the location of the M82 nucleus \citep{lester90}. Negative distances are toward the South, and positive distances are toward the North. Filled circles represent measurements, upside-down triangles are upper-limits, and open squares are fixed to solar values because associated lines are not well constrained. Abundances are relative to solar by number, and error bars are 90\% confidence intervals.}
\vspace{3mm}
\label{fig:gradients}
\end{figure*} 

The outflow spectra show emission lines from O, Ne, Fe L, Mg, Si, and S. The spectra from region S1 has the strongest signal, even moreso than region D, but S1 still lacks the Fe {\sc xxv} line. Regions N1, D, and S1 appear to have significant absorption, precluding the detection of soft X-ray lines, e.g. O~{\sc viii}. However, at larger $r$, the absorption diminishes, and an O~{\sc viii} feature at $\approx$0.65~keV is apparent, particularly in the northern regions. The signal decreases going outward from the major axis, and the hard X-rays (above 2~keV) are less significantly detected in the outer regions, especially S5 and N5. 

Spectra were fit jointly using the model described in Section~\ref{sec:data}, and the results are listed in Table~\ref{table:fitresults}. The fits yielded reduced $\chi^{2}$ values of $0.97-1.62$ with $700-2100$ degrees of freedom (d.o.f. = number of data bins $-$ number of free parameters). Figure~\ref{fig:gradients} shows the best-fit parameters for the 11 regions plotted as a function of $r$, the distance from the starbursting ridge. We find $N_{\rm H}^{\rm M82}$ and the warm-hot $T_{1}$ and hot $T_{2}$ temperature components peak in the center and are lower at greater distances to the north and the south. 

The Si and S abundances are also elevated in the central regions and level out to solar or slightly sub-solar values $\sim$1~kpc from the starburst. Mg and Fe have relatively flat profiles, with median abundances of 1.0~Mg$_{\sun}$ and 0.39~Fe$_{\sun}$, respectively. The O and Ne abundances are uncertain and have larger error bars arising from the associated soft X-ray lines being attenuated by $N_{\rm H}^{\rm M82}$ and the Ne features being blended with Fe L lines. Thus, it is challenging to discern the trends in the O and Ne abundances, but they may be constant. As seen in Figure~\ref{fig:spectra}, prominent O~{\sc viii} lines are apparent in the northern outflow spectra but absent in the southern spectra. Based on our best-fit models, we deduce that this disparity arises because of temperature differences in the {\sc apec} components that lead to less thermal continuum and greater equivalent widths in the O~{\sc viii} in the northern outflow compared to the south.

As noted in Section~\ref{sec:data}, we included a power-law component with $\Gamma=1.5$ in our spectral fits of the 11 regions. Without a power-law component, the fits produced substantial residuals above 3~keV. In the outflow regions (S1$-$S5 and N1$-$N5), the addition of a third thermal component to our fits (rather than a power-law) yielded unphysical best-fit temperatures of $kT_{3} = 10-70$~keV. Moreover, F-tests favored a power-law component over a very hot thermal component with $>$99.5\% confidence. In the case of region D, once the very hot component with $kT_{3}=6.89^{+1.83}_{-0.86}$~keV was included, the power-law improved the fit with 96\% confidence. We note that a steeper power-law with $\Gamma = 2.5-3$ did not adequately fit the data and yielded large residuals. 

\begin{figure*}
\begin{center}
\includegraphics[width=\textwidth]{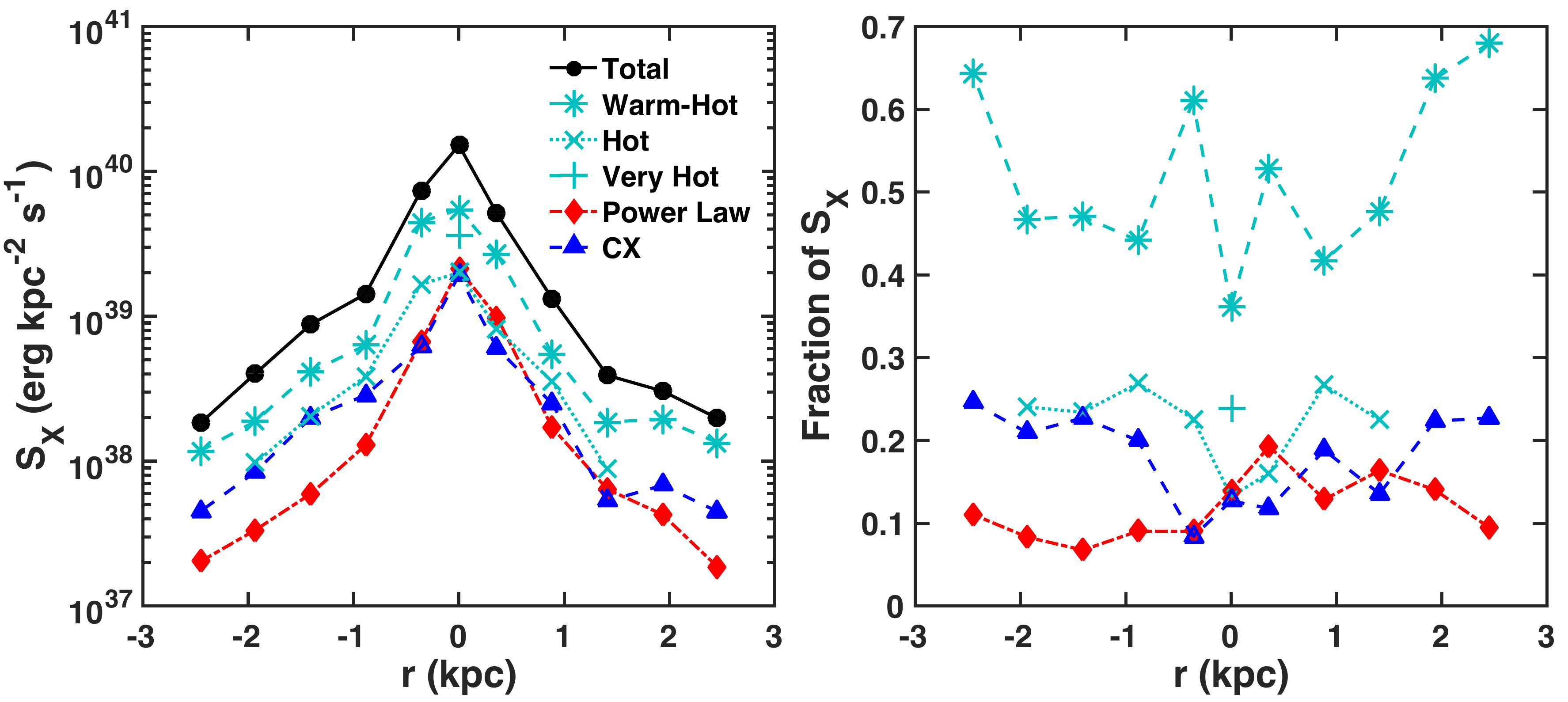}
\end{center}
\vspace{-3mm}
\caption{{\bf Left}: Broad-band ($0.5-7.0$~keV) X-ray surface brightness $S_{\rm X}$ profile of the spectral model components along the M82 minor axis. $S_{\rm X}$ is computed as the absorption-corrected luminosity of each component divided by the area of each region: 0.66~kpc$^{2}$ for region D and 1.65~kpc$^{2}$ for the outflow regions. {\bf Right}: Fraction of the total $S_{\rm X}$ contributed by the spectral model components. The warm-hot component dominates, producing $\approx36-68$\% of the total $S_{\rm X}$, and the other components produce $\approx7-27$\% of the total $S_{\rm X}$.}
\label{fig:Sx}
\end{figure*} 

To examine the relative contribution of the spectral model components to the emission, we plot the broad-band ($0.5-7.0$~keV) X-ray surface brightness $S_{\rm X}$ as a function of $r$ in Figure~\ref{fig:Sx} (left panel). To compute $S_{\rm X}$, we measured the absorption-corrected X-ray luminosity $L_{\rm X}$ of each component in the best spectral fits, and we divided by the area of the 11~regions. The total $S_{\rm X}$ and all of the individual components peak centrally and decline with increasing $r$. Figure~\ref{fig:Sx} (right panel) shows the relative contribution of the model components. The warm-hot plasma is the dominant component, producing $\approx36-68$\% of the total $S_{\rm X}$, whereas the other components produce $\approx7-27$\% of the total $S_{\rm X}$.

In particular, we find that the CX component contributes $8-25$\% of the total broad-band ($0.5-7.0$~keV) absorption-corrected flux (see Figure~\ref{fig:Sx}). The median value is $\approx$20\%, and the central regions (within 500~pc of the starburst) have the smallest CX contribution ($8-13$\%). However, we note that the intrinsic column density $N_{\rm H}^{M82}$ is greatest in these regions, attenuating soft X-rays where CX features are best detected. 

We caution that with many model components and free parameters, the best-fit results may be degenerate. For example, the power-law component is possibly anti-correlated with the hot thermal component, based on their relative contributions to $S_{\rm X}$ as a function of $r$ (as shown in the right panel of Figure~\ref{fig:Sx}). In the future, deeper data are necessary to limit degeneracies in the spectral fits and to better constrain the free parameters.

We estimate the electron number density $n_{\rm e}$ of the thermal plasmas using the best-fit normalizations $norm$ of the {\sc apec} components, since $norm = (10^{-14} EM)/4 \pi D^2$, where $D$ is the distance, $EM = \int n_{\rm e} n_{\rm H} dV$ is the emission measure, and $n_{\rm H}$ is the hydrogen number density. Setting $n_{\rm e} = 1.2 n_{\rm H}$ and integrating over the volume $V$, then $n_{\rm e} = (1.5\times10^{15} norm D^2/fV)^{1/2}$, where $f$ is the filling factor (here, we assume $f = 1$ and caution that $f$ may be lower and not equal between the three hot components). To compute $V$ for each region, we assume a cylindrical geometry and a height equal to the minor axis side of the rectangular regions. To estimate the radius of the cylinder $R_{\rm cyl}$, we produced surface-brightness profiles of the broad-band ($0.5-7$~keV) diffuse emission for each region using 2\arcsec\ slits along the major axis. We define $R_{\rm cyl}$ as half of the size that enclosed 99\% of the emission.  

\begin{deluxetable*}{lcccclccccc}
\tablecolumns{11}
\tablewidth{0pt} \tablecaption{Physical Parameters of the Disk and Outflow Regions\tablenotemark{a} \label{table:ne}} 
\tablehead{\colhead{Reg.} & \colhead{$norm_{1}$\tablenotemark{b}} & \colhead{$norm_{2}$\tablenotemark{b}} & \colhead{$R_{\rm cyl}$} & \colhead{$V$} & \colhead{$n_{\rm e,1}$\tablenotemark{c}} & \colhead{$n_{\rm e,2}$\tablenotemark{c}} & \colhead{$P_{1}/k$\tablenotemark{c}} & \colhead{$P_{2}/k$\tablenotemark{c}} & \colhead{$t_{\rm cool,1}$\tablenotemark{c}} & \colhead{$t_{\rm cool,2}$\tablenotemark{c}} \\
\colhead{} & \colhead{} & \colhead{} & \colhead{($\times10^{21}$ cm)} & \colhead{($\times10^{64}$~cm$^{-3}$)} & \colhead{(cm$^{-3}$)} & \colhead{(cm$^{-3}$)} & \colhead{(K~cm$^{-3}$)} & \colhead{(K~cm$^{-3}$)} & \colhead{(Myr)} & \colhead{(Myr)}}  
\startdata
N5 & 2.3$\times10^{-4}$ & -- & 4.8 & 11.9 & 1.9$\times10^{-2}$ & -- & 1.6$\times10^{5}$ & -- & 96 & -- \\
N4 & 4.5$\times10^{-4}$ & -- & 4.8 & 11.9 & 2.6$\times10^{-2}$ & -- & 2.3$\times10^{5}$ & -- & 68 & -- \\
N3 & 1.9$\times10^{-4}$ & 8.0$\times10^{-5}$ & 4.8 & 11.7 & 1.7$\times10^{-2}$ & 1.1$\times10^{-2}$ & 1.5$\times10^{5}$ & 2.3$\times10^{5}$ & 110 & 368 \\
N2 & 6.3$\times10^{-4}$ & 3.4$\times10^{-4}$ & 4.7 & 11.1 & 3.2$\times10^{-2}$ & 2.4$\times10^{-2}$ & 3.2$\times10^{5}$ & 4.3$\times10^{5}$ & 66 & 154 \\
N1 & 2.0$\times10^{-3}$ & 7.3$\times10^{-4}$ & 4.4 & 9.7 & 6.2$\times10^{-2}$ & 3.7$\times10^{-2}$ & 9.4$\times10^{5}$ & 1.5$\times10^{6}$ & 60 & 393 \\
D & 1.7$\times10^{-3}$ & 7.5$\times10^{-4}$ & 3.7 & 2.7 & 1.1$\times10^{-1}$ & 7.2$\times10^{-2}$ & 1.8$\times10^{6}$ & 2.4$\times10^{6}$ & 32 & 154 \\
S1 & 3.7$\times10^{-3}$ & 1.4$\times10^{-3}$ & 3.8 & 7.4 & 9.6$\times10^{-2}$ & 5.9$\times10^{-2}$ & 1.4$\times10^{6}$ & 1.3$\times10^{6}$ & 30 & 87 \\ 
S2 & 6.8$\times10^{-4}$ & 3.6$\times10^{-4}$ & 4.5 & 10.1 & 3.5$\times10^{-2}$ & 2.6$\times10^{-2}$ & 3.9$\times10^{5}$ & 5.4$\times10^{5}$ & 69 & 178 \\
S3 & 5.2$\times10^{-4}$ & 2.1$\times10^{-4}$ & 4.5 & 10.1 & 3.1$\times10^{-2}$ & 2.0$\times10^{-2}$ & 3.2$\times10^{5}$ & 4.0$\times10^{5}$ & 74 & 222 \\
S4 & 2.3$\times10^{-4}$ & 1.0$\times10^{-4}$ & 4.7 & 11.1 & 2.0$\times10^{-2}$ & 1.3$\times10^{-2}$ & 2.1$\times10^{5}$ & 2.6$\times10^{5}$ & 118 & 322 \\
S5 & 5.4$\times10^{-4}$ & -- & 4.8 & 11.7 & 2.9$\times10^{-2}$ & -- & 2.3$\times10^{5}$ & -- & 65 & -- \\
\enddata
\tablenotetext{a}{The parameters of the very hot component $kT_{3}$ in region D are: $norm_{3} = 1.3\times10^{-3}$, $n_{\rm e,3} = 9.4\times10^{-2}$~cm$^{-3}$, $P_{3}/k = 1.5\times10^{7}$~K~cm$^{-3}$, and $t_{\rm cool,3} = 485$~Myr.}
\tablenotetext{b}{The normalizations are defined as $norm = (10^{-14} EM)/4 \pi D^2$, where $EM  = \int n_{\rm e} n_{\rm H} dV$. Thus, columns 2 and 3 are in units of 10$^{-14}$ cm$^{-5}$.}
\tablenotetext{c}{These parameters were calculated assuming a filling factor $f = 1$. We note that the densities $n_{\rm e} \propto f^{-1/2}$, pressures $P/k \propto f^{-1/2}$, and cooling times $t_{\rm cool} \propto f^{1/2}$.}
\vspace{-5mm}
\end{deluxetable*}

In Table~\ref{table:ne}, we list the best-fit $norm_{1}$, $norm_{2}$, and $norm_{3}$ of the warm-hot, hot, and very hot components, respectively, as well as $R_{\rm cyl}$, $V$, $n_{\rm e,1}$, $n_{\rm e,2}$, and $n_{\rm e,3}$. We find that the densities peak in region D (with $n_{\rm e,1} = 1.1\times10^{-1}$~cm$^{-3}$ and $n_{\rm e,2} = 7.2\times10^{-2}$ there) and fall with $r$. We also compute the thermal pressure $P = 2 n_{\rm e} k T$ and the radiative cooling time $t_{\rm cool} = 3 k T / \Lambda n_{\rm e}$ of the three components, as listed in Table~\ref{table:ne}. $\Lambda$ is the radiative cooling function in units of erg~s$^{-1}$~cm$^{3}$, and we calculate $\Lambda$ at solar metallicity assuming an optically-thin thermal plasma in CIE (as in Figure~1 of \citealt{rosen14} using {\sc chianti}; \citealt{dere97}). 

The warm-hot and hot thermal pressure $P_{1}/k$ and $P_{2}/k$ peak in region D, and these quantities decrease along the minor axis. The very hot thermal pressure $P_{3}/k$ is the dominant term in region D compared to the warm-hot and hot components, with $P_{3}/k = 1.5\times10^{7}$~K~cm$^{-3}$. This value is consistent with the estimates from past X-ray studies \citep{bregman95,strickland09} as well as the central pressure measurements of the ionized gas from optical spectroscopy \citep{heckman90,smith06,westmoquette07}. 

Regions D and S1 have the shortest $t_{\rm cool, 1}$ of $\sim$30~Myr, and the timescale increases to $\sim$100~Myr $\approx$2~kpc outside of the starburst. $t_{\rm cool,2}$ is lowest in region S1 at $\approx$87~Myr and approaches $\sim$400~Myr in the outer regions; $t_{\rm cool,3}$ in region D is the longest of all of the components, with $t_{\rm cool,3} \approx485$ ~Myr. In all regions, $t_{\rm cool}$ of the three components is longer than the advection timescale $t_{\rm adv} \sim r / v$ (where $v$ is the velocity) in the models discussed in Section~\ref{sec:models}, indicating the winds are not radiative.

\section{Discussion} \label{sec:discussion}

\subsection{Comparison to Previous Work} \label{sec:previouswork}

A similar analysis was conducted previously by \cite{ranalli08} who analyzed a 73-ks {\it XMM-Newton} observation of M82 and measured how the temperature and abundance pattern varied $\lesssim$2.5~kpc from the major axis. They extracted spectra from 5 regions north and south of the M82 galaxy plane, much like the setup in this paper, though their central region was larger than our region D and had a circular aperture with a 1\arcmin\ diameter. They fit a two-temperature plasma model and found a very hot component of $\sim$7~keV $\approx8\times10^{7}$~K that was relatively constant along the minor axis. Their warm-hot component decreased from 0.53~keV ($\approx6\times10^{6}$~K) to 0.35~keV ($\approx4\times10^{6}$~K) from the center to their outer regions. \cite{ranalli08} found that several elements (O, Ne, Mg, and Fe) were substantially more abundant in the outflows than in the disk: e.g., O and Ne were $\sim$10$\times$ more abundant $\sim$2~kpc away in the outflows than in the central region. This trend was less pronounced for Si (with a $\sim$2$\times$ enhancement in the outflows relative to the disk), whereas S was most abundant in the central regions (though large errors precluded reliable measurements in the outer regions). 

In addition to the two thermal plasmas in their model of the central region's spectra, \cite{ranalli08} included a power-law component (with spectral index $\Gamma = 1.60^{+0.04}_{-0.03}$) to account for point sources as well as Gaussian functions at 0.78~keV and 1.2~keV to represent two spectral lines associated with CX (we note that a physical model of CX emission was not publicly available when that work was completed). \cite{ranalli08} did not include the power-law component in their fits to the outflow spectra, based on the reasoning that point sources were resolved sufficiently to be excluded from the extraction regions.

We find a similar warm-hot temperature profile and very hot temperature in the starburst as \cite{ranalli08}, but the abundance profiles between the two works are different. Thus, it is worth considering why these discrepancies have arisen. In contrast to \cite{ranalli08}, we have included a CX and power-law component in the outflow spectral models in this work. The former is necessary for reliable measurements of O, Ne, and Mg abundances as the CX flux contributes $\sim$50\% to the emission lines associated with these elements in M82 \citep{liu11}. Furthermore, as Fe L features are blended with Ne~{\sc ix} and Ne~{\sc x} at energies $\approx0.9 - 1.2$~keV, the measured Fe abundance is also be affected by not including a CX component. Moreover, we note that \cite{zhang14} measured abundances from fits to {\it XMM-Newton} RGS data using an absorbed {\sc apec}$+$CX model, and their values were consistent (within the errors) of our region~D values. Thus, the different abundance profiles (particularly O, Ne, Mg, and Fe) from this work and \cite{ranalli08} likely arise from our inclusion of a CX component.

\begin{figure*}
\begin{center}
\includegraphics[width=\textwidth]{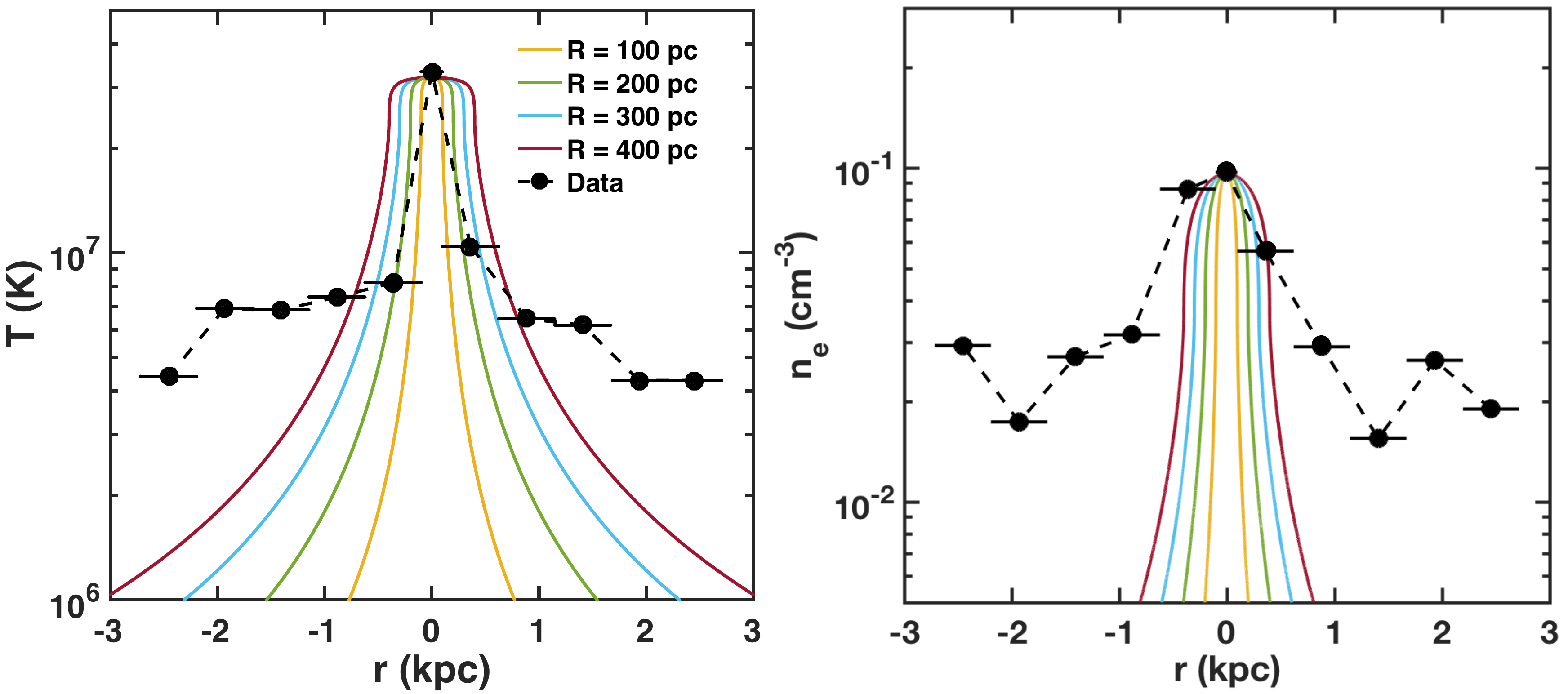}
\end{center}
\vspace{-3mm}
\caption{Flux-weighted hot-gas temperature $T$ (left) and density $n_{\rm e}$ (right) profile of M82 (black lines) compared to four adiabatic model predictions (solid lines) with $R = [100, 200, 300, 400]$~pc, $\alpha = [0.02, 0.08, 0.17, 0.31]$, and $\beta = [0.01, 0.03, 0.08, 0.14]$, respectively. In the adiabatic scenarios, $T \propto r^{-4/3}$ and $n_{\rm e} \propto r^{-2}$, and these profiles are steeper than are observed and may suggest mass loading in the hot superwind.}
\label{fig:modelcomparison}
\end{figure*}  

Unlike \cite{ranalli08}, we only find the very hot component (with $kT_{3} = 6.89^{+1.83}_{-0.86}$~keV) in the M82 central region, and F-tests show with $>$99.5\% confidence that it is not present in any of the outflow regions. Our result is consistent with the maps produced by \cite{strickland07} who showed that the Fe~{\sc xxv} (associated with the very hot component) only extends $<$100~pc along the M82 minor axis. A likely explanation for these discrepant results is that our power-law component (which was not included in the Ranalli et al. fits) accounted for the hard X-ray flux such that a very hot component was not necessary. F-tests confirmed that a power-law with photon index $\Gamma = 1.5$ was preferred over the inclusion of a very hot thermal component in all outflow regions analyzed in this work. Moreover, the lack of a Fe~{\sc xxv} line in the outflow spectra (see the right panels of Figure~\ref{fig:spectra}) -- which was also noted by \cite{ranalli08} from their {\it XMM-Newton} outflow spectra -- reinforce the possibility that the hard X-ray emission is non-thermal in nature.

The other previous study that explored how the hot gas temperature and abundance profiles vary in the M82 outflows was conducted by \cite{konami11} using 101 ks of {\it Suzaku} observations. They reported that two- or three-temperature plasma components (including a soft $\sim$0.25~keV temperature plasma) were necessary to describe the $0.5-6$~keV spectra of the M82 disk and three outflow regions north of the disk. Our warm-hot and hot temperature profiles are consistent with the \cite{konami11} results. They did not find a very hot component in their disk region, but it is likely because the bandpass they considered did not include the Fe~{\sc xxv} line. Additionally, \cite{konami11} found no spatial variation in the abundance ratios (O/Fe, Ne/Fe, Mg/Fe) of the hot plasmas, similar to our results. 

\subsection{Comparison to Wind Models} \label{sec:models}

The profiles presented in Section~\ref{sec:results} can be compared to superwind model predictions to constrain outflow properties. \cite{chevalier85} developed a simple spherically-symmetric wind model to explain the extended X-ray emission of M82, whereby SNe inject mass and energy at rates $\dot{M}$ and $\dot{E}$, respectively, in a region of size $R$. The energy injection rate is $\dot{E} = \alpha \dot{E}_{\rm SN}$ (where $\dot{E}_{\rm SN}$ is the energy injection from SNe, $\sim10^{51}$~erg per 100~$M_{\sun}$ of star formation), and the mass injection rate is $\dot{M} = \beta \dot{M_{\ast}}$ (where $\dot{M_{\ast}}$ is the star formation rate). By energy conservation (neglecting radiative cooling and gravity; see \citealt{thompson16} and references therein for the effects of radiative cooling), the asymptotic velocity of the $v_{\rm hot, \infty} = (2 \dot{E}/\dot{M})^{1/2} \simeq 10^{3} (\alpha/\beta)^{1/2}$~km~s$^{-1}$ and the temperature at $r = R$ is $T_{\rm hot} = (m_{\rm p}/k)(3/20)v_{\rm hot,\infty}^{2} \simeq 2\times10^{7} (\alpha/\beta)$~K. Although the CC85 model assumes a steady-state wind, this assumption is physically justified by the short sound crossing time ($\lesssim$10~Myr) across the starburst region for the hot phase, and the CC85 model is a useful comparison point (e.g., \citealt{strickland09}).

Based on their observational measurements from hard X-ray lines detected in the central 500\,pc of M82, \cite{strickland09} found a central temperature of $T_{\rm c} = (3-8)\times10^{7}$~K, central density of $n_{\rm c} \sim 0.2$~cm$^{-3}$, and central pressure of $P_{\rm c}/k = (1-3)\times10^{7}$~K~cm$^{-3}$. From these values, assuming an injection region of $R=300$\,pc and $\dot{M_{\ast}} = 6$~$M_{\sun}$ yr$^{-1}$, they estimated $0.5 \leq \alpha \leq 2.4$ and $0.2 \leq \beta \leq 0.6$.\footnote{Note that we have converted \cite{strickland09}'s estimates using our definitions of the mass loading and thermalization parameters for consistency.} Our results for the best-fit parameters associated with the very hot component in region~D are consistent with these previous findings. Assuming $\dot{M_{\ast}} = 6$~$M_{\sun}$ yr$^{-1}$ and an injection region of $R =[100,200,300,400]$~pc, fitting only the very hot component in region D, we find that $\alpha = [0.07,0.26,0.59,1.0]$, and $\beta = [0.01,0.05,0.11,0.19]$, respectively. For a fiducial value of $R=300$\,pc, these numbers reflect relatively high values of the thermalization efficiency and low values of the mass loading. 

The numbers change significantly if we instead fit to the flux-weighted values of the temperature and density in the core. Figure~\ref{fig:modelcomparison} shows the flux-weighted temperature $T$ and density $n_{\rm e}$ profiles of our data and of four \cite{chevalier85} models tuned to the central values with $\alpha = [0.02,0.08,0.17,0.31]$, and $\beta = [0.01,0.03,0.08,0.14]$, respectively. These lower values of $\alpha$ and $\beta$ relative to those derived from only the hottest component in the core result from the fact that the warm-hot component dominates in the flux-weighted temperature and density (see Figure \ref{fig:Sx}).

Outside this central region (for radii $r>R$), the \cite{chevalier85} model assumes that the outflowing gas experiences adiabatic expansion such that $T \propto r^{-4/3}$ (for an adiabatic index $\gamma$ = 5/3), $n \propto r^{-2}$, and pressure $P \propto r^{-10/3}$ once the asymptotic velocity of the wind has been reached. Near the core region $R$, where the flow accelerates, the predicted profiles are steeper. As shown in Figure \ref{fig:modelcomparison}, the data diverge substantially from the models $>$1~kpc from the midplane: our flux-weighted temperature and density profiles are much flatter than predicted by spherical adiabatic expansion.

One possible reason is that mass loading is occurring in the hot superwind (e.g., \citealt{suchkov94,suchkov96}). Cool gas clouds driven out of galactic disks by hot winds are quickly shredded by hydrodynamical instabilities (e.g., \citealt{sca15,zhang17}), and this material mixes and enters the hot phase. Recently, \cite{schneider20} examined the interplay between the hot and cold components in galactic winds, showing that mixing between the hot and cooler phase leads to shallower density and temperature profiles that are qualitatively similar to those in Figure~\ref{fig:modelcomparison}. In the future, flexible models of hot gas entrainment and of non-spherical areal divergence are needed to assess the possibility that this physics explains the temperature and density gradients derived from the X-ray observations (e.g., \citealt{suchkov96}). Such a comparison between models and data is critical to constrain the energy thermalization and mass loading in M82's superwind.

\subsection{Metal Abundance Profiles} 

As shown in Figure~\ref{fig:gradients} and discussed in Section~\ref{sec:results}, the O, Ne, Mg, and Fe abundances are relatively constant along the minor axis. While the O and Mg abundances are solar metallicity, Ne is about 1.5$\times$ solar. These values are consistent with the abundances measured in M82 stars and H{\sc ii} regions \citep{achtermann95,origlia04}. The Si and S are enhanced to 2.5$\times$ and 3.5$\times$ solar, respectively, in the central 500~pc, indicative of enrichment from SN ejecta there. As described in Section~\ref{sec:data}, to limit free parameters in our spectral fits, we tied the abundances between the warm-hot and hot components, yielding the temperature-weighted abundance gradients plotted in Figure~\ref{fig:gradients}. We note that the Si and S line fluxes originates predominantly from the hot component, whereas the O, Ne, Mg, and Fe L line fluxes arise from the warm-hot component. Thus, the radial trend observed for Si and S may reflect the temperature gradient and enrichment of the hot phase.

Three-dimensional hydrodynamical simulations of superwinds by \cite{melioli13} predict that the hot gas component can be metal-rich with average metallicities of $\bar{Z} \sim4.5 Z_{\sun}$, whereas the colder, denser material will maintain solar metallicity. \cite{melioli13} suggested that a large fraction of the ejected metals is retained in and around the galactic disk, with $\sim$1/3 of those metals transported to enrich the intergalactic medium (IGM). Our results are consistent with these predictions, showing that the SN-rich ejecta only extends $\approx$500~pc from the starbursting ridge.

The stellar disks of spiral galaxies are believed to be surrounded by the CGM out to the virial radius (e.g. \citealt{oppenheimer16}). The CGM, which is believed to be the largest baryon reservoir in a galaxy, is dominated by warm-hot gas of temperatures of $T \sim10^{6}$~K in $L^{\ast}$ galaxies, with $T \propto M_{\rm halo}^{2/3}$ where $M_{\rm halo}$ is the halo mass. Highly ionized metal lines are detected from the warm-hot CGM, indicating a metal-rich CGM (e.g., \citealt{gupta12} and references therein) that originates from the stellar disk and is ejected in galactic outflows. The observations reported here show that metals are being expelled in these superwinds. Our results are consistent with the super-solar Ne/O and Ne/Fe ratios detected in the Milky Way CGM  \citep{das19a,das19b} and support the direct connection between the wind and the CGM.

\subsection{Charge Exchange}

As described in Section~\ref{sec:data}, we include a CX component in our spectral fits that contributes $8-25$\% of the total broad-band flux (see Figure~\ref{fig:Sx}) in our 11 regions. Previous work using high-resolution X-ray spectroscopy has measured the CX contribution to M82's X-ray emission overall. Using {\it XMM-Newton} RGS data, \cite{ranalli08} found that the O lines could not be accounted for properly with a multi-temperature thermal plasma model, suggesting a significant flux from CX. \cite{liu11} quantified the CX contribution to K$\alpha$ triplets of He-like O, Ne, and Mg: 90, 50, and 30\%, respectively. \cite{zhang14} analyzed {\it XMM-Newton} RGS spectra of the central outflow and EPIC-pn spectra of the Cap, adopting a physical model (from \citealt{smith12}) that included one thermal plasma and a CX component. They found that $\approx$25\% of the flux from $0.4-2$~keV arises from CX. 

Our results are consistent with these prior studies of M82 and show that the CX emission contributes non-negligibly to at least $\pm$2.5~kpc from the starburst. CX occurs at the interface between hot and cold gases, and as noted by \cite{zhang14}, the CX in M82 may be a by-product of mass loading. Consequently, the finding of CX emission along the M82 minor axis suggests that the mixing and heating of cool gas into the warm-hot phase persists out to large distances.

\subsection{Non-thermal Emission} 

\begin{figure}[t]
\begin{center}
\includegraphics[width=0.8\columnwidth]{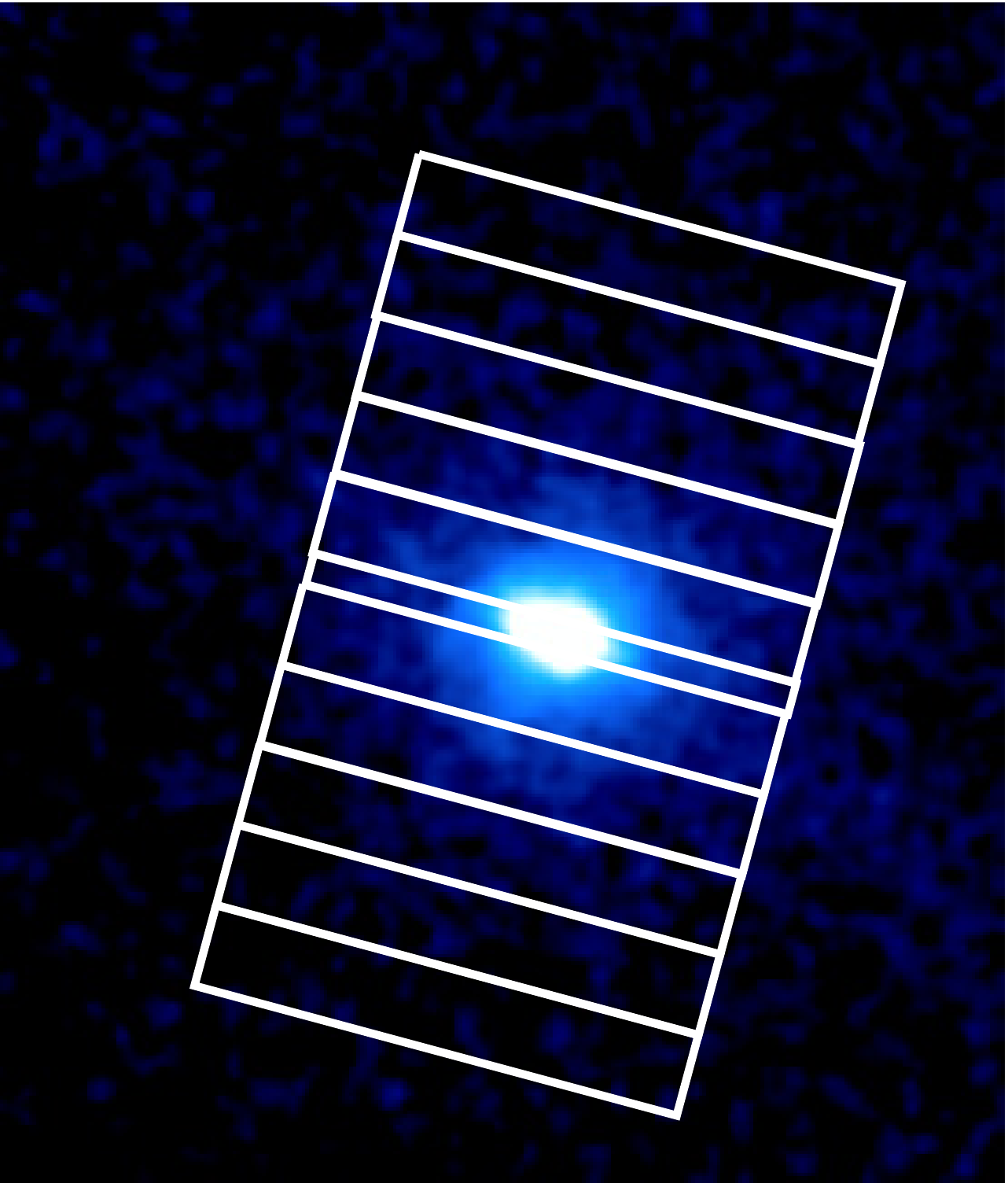}
\end{center}
\vspace{-3mm}
\caption{Exposure-corrected image of the diffuse hard X-ray emission in the $4-6$~keV band. In this energy band, the power-law produces $80-100$\% of the flux in the outflow regions and 40\% of the flux in region D. The image has been smoothed with a Gaussian of width $\sigma$ = 3 pixels. White boxes denote the 11 regions where spectra were extracted (see Figure~\ref{fig:spectra}). The emission peaks in the starburst ridge and falls off with distance along the minor axis. North is up, and East is left.}
\label{fig:hardX}
\end{figure}  

As described in Section~\ref{sec:results}, a power-law component is necessary to adequately fit the spectra in all 11 regions. The combined luminosity of the 11 regions from the power-law component is $L_{\rm X,PL} = (5.0\pm0.3)\times10^{39}$ erg~s$^{-1}$, $\approx$13\% of the total $L_{\rm X}$ in the $0.5-7$~keV band. To demonstrate the distribution of the associated emission, Figure~\ref{fig:hardX} is an exposure-corrected image of the diffuse hard ($4-6$~keV) X-rays, an energy band where the power-law component produces $80-100$\% of the flux in the outflow regions and 40\% of the flux in region D. The hard X-ray emission peaks in region D and fall off with distance along the minor axis. It is likely that some fraction of this emission, particularly in the center of M82, arises from unresolved X-ray binaries, so $L_{\rm X, PL}$ should be viewed as upper-limit on a diffuse non-thermal component. 

The presence and nature of diffuse hard X-ray emission in M82 is debated. Using the first {\it Chandra} observations of M82, \cite{griffiths00} attributed the diffuse hard X-rays detected in the nuclear region to thermal bremsstrahlung from a $\sim4\times10^{7}$~K plasma because of the detected Fe~{\sc xxv} line. \cite{strickland07} also examined the diffuse hard X-rays in the central 500~pc of M82 using {\it Chandra} and {\it XMM-Newton} data. They concluded that $20-30$\% of the emission was truly diffuse and that the continuum was better fit as a power-law with a photon index of $\Gamma=2.5-3$ than with thermal bremsstrahlung models. In their analysis of deeper {\it XMM-Newton} observations (as described in Section~\ref{sec:previouswork}), \cite{ranalli08} used a power-law component to account for unresolved X-ray binaries, but this component was not included in their models of the outflow region spectra. 

While some of the emission may arise from point sources, the power-law component in the outflows may arise from diffuse non-thermal X-rays. Past studies have proposed that inverse-Compton (IC) scattering of IR photons by relativistic electrons may produce significant diffuse hard X-rays in M82 \citep{schaaf89,moran97}. In this scenario, the intense IR emission of the starburst up-scatters target photons to $\sim$100~MeV energies, and the spectrum extends down to X-ray energies with a hard photon index of $\Gamma \approx 1.0-1.5$. \cite{lacki13} suggest that the non-thermal diffuse hard X-ray emission is produced by synchrotron. Although with substantial uncertainties, in their fiducial models they find that IC and synchrotron are able to produce only $\sim1$\% and $\sim2$\%, respectively, of the observed diffuse hard X-rays. Recent {\sc galprop} models from \cite{buckman20} predict that $\approx66-90$\% of the non-thermal core and halo emission is from IC in the $1-10$~keV band, but the total luminosity falls short of that observed by a factor of order $\sim10$.

In the future, more X-ray observations of the outflow regions, particularly with hard X-ray sensitivity and sub-arcsecond spatial resolution to disentangle point sources, are necessary to verify the presence and nature of the non-thermal emission. 

\section{Conclusions} \label{sec:conclusions}

We analyze deep {\it Chandra} observations of M82 to produce images and spectra of the diffuse X-ray emission of the starburst and outflows. Based on fits to the spectra from 11 regions up to $\pm$2.5~kpc from the starburst ridge, we find that the intrinsic column densities, the plasma temperatures, gas densities, and the Si and S abundances peak in the M82 center and decrease $>$500~pc away. By contrast, the O, Ne, Mg, and Fe abundances are relatively constant between the starburst and the outflows, indicating effective transport of stellar disk metals to the CGM. We compare the observed temperature and gas density profiles to superwind model predictions, and we show that these profiles are much shallower than expected for adiabatic expansion (see Figure~\ref{fig:modelcomparison}). This result suggests that the hot winds are being mass loaded due to the mixing and heating of cooler gas into the hot phase.

In addition to the thermal components in our spectral models, we find it is necessary to include charge exchange and a power-law component in all of the regions considered in this work. The CX contributes $8-25$\% to the total broad-band flux (consistent with the work of \citealt{zhang14}), with the smallest contribution within $\sim$500~pc of the M82 nucleus. However, the high intrinsic column density there attenuates soft X-rays where CX features are best detected. We also find that a hard power-law component is required throughout the central and outflow regions. This power-law component accounts for $\approx$13\% of the total broad-band flux in the 11 regions, and the spectral fits favored a shallow photon index of $\sim$1.5. The origin of this power-law component is uncertain and may be unresolved point sources, inverse-Compton scattering of IR photons, or synchrotron producing a non-thermal halo. 

In the future, detailed wind models over a wide parameter space of energy thermalization and mass loading are necessary to probe the outflow properties. Observationally, X-ray calorimeters (like on the upcoming {\it XRISM} mission; \citealt{tashiro18}) will facilitate measurements of the hot gas velocity $v_{\rm hot}$, which will provide strong constraints on the energy content and mass loading of the hot wind. When calorimeters achieve arcsecond spatial resolution, as in the proposed {\it Lynx} X-ray Observatory \citep{lynx18}, they will enable measurements of the metal abundances and velocities along the minor axes of starburst-driven winds in many galaxies. Hard X-ray sensitivity (at $\sim$6~keV) is crucial for characterization of the Fe~{\sc xxv} in the starburst cores, and soft X-ray capabilities (below 1~keV) are vital to probe the interplay of warm-hot and cooler gas in the galaxy outflows \citep{hodges19}.

\acknowledgements

We thank the anonymous reviewer for constructive feedback that improved the manuscript. We also thank Adam Leroy, Paul Martini, David Weinberg, Patrick Slane, and the OSU Galaxy/ISM Meeting for useful discussions. LAL is supported by a Cottrell Scholar Award from the Research Corporation of Science Advancement. DDN and TAT are supported in part by National Science Foundation Grant \#1516967 and NASA ATP 80NSSC18K0526. 

\software{CIAO (v4.7; \citealt{fru06}), XSPEC (v12.9.0; \citealt{arnaud96})}

\nocite{*}
\bibliographystyle{aasjournal}
\bibliography{m82bib.bib}

\begin{thebibliography}{}
\expandafter\ifx\csname natexlab\endcsname\relax\def\natexlab#1{#1}\fi
\providecommand{\url}[1]{\href{#1}{#1}}
\providecommand{\dodoi}[1]{doi:~\href{http://doi.org/#1}{\nolinkurl{#1}}}
\providecommand{\doeprint}[1]{\href{http://ascl.net/#1}{\nolinkurl{http://ascl.net/#1}}}
\providecommand{\doarXiv}[1]{\href{https://arxiv.org/abs/#1}{\nolinkurl{https://arxiv.org/abs/#1}}}

\bibitem[{{Achtermann} \& {Lacy}(1995)}]{achtermann95}
{Achtermann}, J.~M., \& {Lacy}, J.~H. 1995, \apj, 439, 163,
  \dodoi{10.1086/175161}

\bibitem[{{Arnaud}(1996)}]{arnaud96}
{Arnaud}, K.~A. 1996, Astronomical Society of the Pacific Conference Series,
  Vol. 101, {XSPEC: The First Ten Years}, ed. G.~H. {Jacoby} \& J.~{Barnes}, 17

\bibitem[{{Asplund} {et~al.}(2009){Asplund}, {Grevesse}, {Sauval}, \&
  {Scott}}]{asplund09}
{Asplund}, M., {Grevesse}, N., {Sauval}, A.~J., \& {Scott}, P. 2009, \araa, 47,
  481, \dodoi{10.1146/annurev.astro.46.060407.145222}

\bibitem[{{Beir{\~a}o} {et~al.}(2015){Beir{\~a}o}, {Armus}, {Lehnert},
  {Guillard}, {Heckman}, {Draine}, {Hollenbach}, {Walter}, {Sheth}, {Smith},
  {Shopbell}, {Boulanger}, {Surace}, {Hoopes}, \& {Engelbracht}}]{beirao15}
{Beir{\~a}o}, P., {Armus}, L., {Lehnert}, M.~D., {et~al.} 2015, \mnras, 451,
  2640, \dodoi{10.1093/mnras/stv1101}

\bibitem[{{Booth} {et~al.}(2013){Booth}, {Agertz}, {Kravtsov}, \&
  {Gnedin}}]{booth13}
{Booth}, C.~M., {Agertz}, O., {Kravtsov}, A.~V., \& {Gnedin}, N.~Y. 2013,
  \apjl, 777, L16, \dodoi{10.1088/2041-8205/777/1/L16}

\bibitem[{{Borthakur} {et~al.}(2013){Borthakur}, {Heckman}, {Strickland},
  {Wild}, \& {Schiminovich}}]{bor13}
{Borthakur}, S., {Heckman}, T., {Strickland}, D., {Wild}, V., \&
  {Schiminovich}, D. 2013, \apj, 768, 18, \dodoi{10.1088/0004-637X/768/1/18}

\bibitem[{{Bregman} {et~al.}(1995){Bregman}, {Schulman}, \&
  {Tomisaka}}]{bregman95}
{Bregman}, J.~N., {Schulman}, E., \& {Tomisaka}, K. 1995, \apj, 439, 155,
  \dodoi{10.1086/175160}

\bibitem[{{Brightman} {et~al.}(2016){Brightman}, {Harrison}, {Walton},
  {Fuerst}, {Hornschemeier}, {Zezas}, {Bachetti}, {Grefenstette}, {Ptak},
  {Tendulkar}, \& {Yukita}}]{brightman16}
{Brightman}, M., {Harrison}, F., {Walton}, D.~J., {et~al.} 2016, \apj, 816, 60,
  \dodoi{10.3847/0004-637X/816/2/60}

\bibitem[{{Buckman} {et~al.}(2020){Buckman}, {Linden}, \&
  {Thompson}}]{buckman20}
{Buckman}, B.~J., {Linden}, T., \& {Thompson}, T.~A. 2020, \mnras, 494, 2679,
  \dodoi{10.1093/mnras/staa875}

\bibitem[{{Cappi} {et~al.}(1999){Cappi}, {Persic}, {Bassani}, {Franceschini},
  {Hunt}, {Molendi}, {Palazzi}, {Palumbo}, {Rephaeli}, \& {Salucci}}]{cappi99}
{Cappi}, M., {Persic}, M., {Bassani}, L., {et~al.} 1999, \aap, 350, 777.
\newblock \doarXiv{astro-ph/9908312}

\bibitem[{{Chevalier} \& {Clegg}(1985)}]{chevalier85}
{Chevalier}, R.~A., \& {Clegg}, A.~W. 1985, \nat, 317, 44,
  \dodoi{10.1038/317044a0}

\bibitem[{{Coker} {et~al.}(2013){Coker}, {Thompson}, \& {Martini}}]{coker13}
{Coker}, C.~T., {Thompson}, T.~A., \& {Martini}, P. 2013, \apj, 778, 79,
  \dodoi{10.1088/0004-637X/778/1/79}

\bibitem[{{Das} {et~al.}(2019{\natexlab{a}}){Das}, {Mathur}, {Gupta},
  {Nicastro}, \& {Krongold}}]{das19a}
{Das}, S., {Mathur}, S., {Gupta}, A., {Nicastro}, F., \& {Krongold}, Y.
  2019{\natexlab{a}}, \apj, 887, 257, \dodoi{10.3847/1538-4357/ab5846}

\bibitem[{{Das} {et~al.}(2019{\natexlab{b}}){Das}, {Mathur}, {Nicastro}, \&
  {Krongold}}]{das19b}
{Das}, S., {Mathur}, S., {Nicastro}, F., \& {Krongold}, Y. 2019{\natexlab{b}},
  \apjl, 882, L23, \dodoi{10.3847/2041-8213/ab3b09}

\bibitem[{{Dere} {et~al.}(1997){Dere}, {Landi}, {Mason}, {Monsignori Fossi}, \&
  {Young}}]{dere97}
{Dere}, K.~P., {Landi}, E., {Mason}, H.~E., {Monsignori Fossi}, B.~C., \&
  {Young}, P.~R. 1997, \aaps, 125, 149, \dodoi{10.1051/aas:1997368}

\bibitem[{{Dickey} \& {Lockman}(1990)}]{dickey90}
{Dickey}, J.~M., \& {Lockman}, F.~J. 1990, \araa, 28, 215,
  \dodoi{10.1146/annurev.aa.28.090190.001243}

\bibitem[{{Engelbracht} {et~al.}(2006){Engelbracht}, {Kundurthy}, {Gordon},
  {Rieke}, {Kennicutt}, {Smith}, {Regan}, {Makovoz}, {Sosey}, {Draine},
  {Helou}, {Armus}, {Calzetti}, {Meyer}, {Bendo}, {Walter}, {Hollenbach},
  {Cannon}, {Murphy}, {Dale}, {Buckalew}, \& {Sheth}}]{engel06}
{Engelbracht}, C.~W., {Kundurthy}, P., {Gordon}, K.~D., {et~al.} 2006, \apjl,
  642, L127, \dodoi{10.1086/504590}

\bibitem[{{Everett} {et~al.}(2008){Everett}, {Zweibel}, {Benjamin}, {McCammon},
  {Rocks}, \& {Gallagher}}]{everett08}
{Everett}, J.~E., {Zweibel}, E.~G., {Benjamin}, R.~A., {et~al.} 2008, \apj,
  674, 258, \dodoi{10.1086/524766}

\bibitem[{{Foster} {et~al.}(2012){Foster}, {Ji}, {Smith}, \&
  {Brickhouse}}]{foster12}
{Foster}, A.~R., {Ji}, L., {Smith}, R.~K., \& {Brickhouse}, N.~S. 2012, \apj,
  756, 128, \dodoi{10.1088/0004-637X/756/2/128}

\bibitem[{{Freedman} {et~al.}(1994){Freedman}, {Hughes}, {Madore}, {Mould},
  {Lee}, {Stetson}, {Kennicutt}, {Turner}, {Ferrarese}, {Ford}, {Graham},
  {Hill}, {Hoessel}, {Huchra}, \& {Illingworth}}]{freedman94}
{Freedman}, W.~L., {Hughes}, S.~M., {Madore}, B.~F., {et~al.} 1994, \apj, 427,
  628, \dodoi{10.1086/174172}

\bibitem[{{Fruscione} {et~al.}(2006){Fruscione}, {McDowell}, {Allen},
  {Brickhouse}, {Burke}, {Davis}, {Durham}, {Elvis}, {Galle}, {Harris},
  {Huenemoerder}, {Houck}, {Ishibashi}, {Karovska}, {Nicastro}, {Noble},
  {Nowak}, {Primini}, {Siemiginowska}, {Smith}, \& {Wise}}]{fru06}
{Fruscione}, A., {McDowell}, J.~C., {Allen}, G.~E., {et~al.} 2006, Society of
  Photo-Optical Instrumentation Engineers (SPIE) Conference Series, Vol. 6270,
  {CIAO: Chandra's data analysis system}, 62701V

\bibitem[{{Gerke} {et~al.}(2011){Gerke}, {Kochanek}, {Prieto}, {Stanek}, \&
  {Macri}}]{gerke11}
{Gerke}, J.~R., {Kochanek}, C.~S., {Prieto}, J.~L., {Stanek}, K.~Z., \&
  {Macri}, L.~M. 2011, \apj, 743, 176, \dodoi{10.1088/0004-637X/743/2/176}

\bibitem[{{Griffiths} {et~al.}(2000){Griffiths}, {Ptak}, {Feigelson},
  {Garmire}, {Townsley}, {Brandt}, {Sambruna}, \& {Bregman}}]{griffiths00}
{Griffiths}, R.~E., {Ptak}, A., {Feigelson}, E.~D., {et~al.} 2000, Science,
  290, 1325, \dodoi{10.1126/science.290.5495.1325}

\bibitem[{{Gupta} {et~al.}(2012){Gupta}, {Mathur}, {Krongold}, {Nicastro}, \&
  {Galeazzi}}]{gupta12}
{Gupta}, A., {Mathur}, S., {Krongold}, Y., {Nicastro}, F., \& {Galeazzi}, M.
  2012, \apjl, 756, L8, \dodoi{10.1088/2041-8205/756/1/L8}

\bibitem[{{Heckman} {et~al.}(1990){Heckman}, {Armus}, \& {Miley}}]{heckman90}
{Heckman}, T.~M., {Armus}, L., \& {Miley}, G.~K. 1990, \apjs, 74, 833,
  \dodoi{10.1086/191522}

\bibitem[{{Heckman} \& {Thompson}(2017)}]{heckman17}
{Heckman}, T.~M., \& {Thompson}, T.~A. 2017, {Galactic Winds and the Role
  Played by Massive Stars}, ed. A.~W. {Alsabti} \& P.~{Murdin}, 2431

\bibitem[{{Hodges-Kluck} {et~al.}(2019){Hodges-Kluck}, {Lopez}, {Yukita},
  {Ptak}, {Swartz}, {Tzanavaris}, {Veilleux}, \& {Bregman}}]{hodges19}
{Hodges-Kluck}, E., {Lopez}, L.~A., {Yukita}, M., {et~al.} 2019, \baas, 51,
  257.
\newblock \doarXiv{1903.09692}

\bibitem[{{Hoopes} {et~al.}(2005){Hoopes}, {Heckman}, {Strickland }, {Seibert},
  {Madore}, {Rich}, {Bianchi}, {Gil de Paz}, {Burgarella}, {Thilker},
  {Friedman}, {Barlow}, {Byun}, {Donas}, {Forster}, {Jelinsky}, {Lee},
  {Malina}, {Martin}, {Milliard}, {Morrissey}, {Neff}, {Schiminovich},
  {Siegmund}, {Small}, {Szalay}, {Welsh}, \& {Wyder}}]{hoopes05}
{Hoopes}, C.~G., {Heckman}, T.~M., {Strickland }, D.~K., {et~al.} 2005, \apjl,
  619, L99, \dodoi{10.1086/423032}

\bibitem[{{Ipavich}(1975)}]{ipavich75}
{Ipavich}, F.~M. 1975, \apj, 196, 107, \dodoi{10.1086/153397}

\bibitem[{{Kaneda} {et~al.}(2010){Kaneda}, {Ishihara}, {Suzuki}, {Ikeda},
  {Onaka}, {Yamagishi}, {Ohyama}, {Wada}, \& {Yasuda}}]{kaneda10}
{Kaneda}, H., {Ishihara}, D., {Suzuki}, T., {et~al.} 2010, \aap, 514, A14,
  \dodoi{10.1051/0004-6361/200913769}

\bibitem[{{Kennicutt} {et~al.}(2008){Kennicutt}, {Lee}, {Funes}, {J.}, {Sakai},
  \& {Akiyama}}]{kenn08}
{Kennicutt}, Robert~C., J., {Lee}, J.~C., {Funes}, J.~G., {et~al.} 2008, \apjs,
  178, 247, \dodoi{10.1086/590058}

\bibitem[{{Konami} {et~al.}(2011){Konami}, {Matsushita}, {Tsuru}, {Gand hi}, \&
  {Tamagawa}}]{konami11}
{Konami}, S., {Matsushita}, K., {Tsuru}, T.~G., {Gand hi}, P., \& {Tamagawa},
  T. 2011, \pasj, 63, S913, \dodoi{10.1093/pasj/63.sp3.S913}

\bibitem[{{Lacki} \& {Thompson}(2013)}]{lacki13}
{Lacki}, B.~C., \& {Thompson}, T.~A. 2013, \apj, 762, 29,
  \dodoi{10.1088/0004-637X/762/1/29}

\bibitem[{{Leeuw} \& {Robson}(2009)}]{leeuw09}
{Leeuw}, L.~L., \& {Robson}, E.~I. 2009, \aj, 137, 517,
  \dodoi{10.1088/0004-6256/137/1/517}

\bibitem[{{Lehnert} {et~al.}(1999){Lehnert}, {Heckman}, \&
  {Weaver}}]{lehnert99}
{Lehnert}, M.~D., {Heckman}, T.~M., \& {Weaver}, K.~A. 1999, \apj, 523, 575,
  \dodoi{10.1086/307762}

\bibitem[{{Leroy} {et~al.}(2015){Leroy}, {Walter}, {Martini}, {Roussel},
  {Sandstrom}, {Ott}, {Weiss}, {Bolatto}, {Schuster}, \&
  {Dessauges-Zavadsky}}]{leroy15}
{Leroy}, A.~K., {Walter}, F., {Martini}, P., {et~al.} 2015, \apj, 814, 83,
  \dodoi{10.1088/0004-637X/814/2/83}

\bibitem[{{Lester} {et~al.}(1990){Lester}, {Carr}, {Joy}, \&
  {Gaffney}}]{lester90}
{Lester}, D.~F., {Carr}, J.~S., {Joy}, M., \& {Gaffney}, N. 1990, \apj, 352,
  544, \dodoi{10.1086/168557}

\bibitem[{{Liu} {et~al.}(2011){Liu}, {Mao}, \& {Wang}}]{liu11}
{Liu}, J., {Mao}, S., \& {Wang}, Q.~D. 2011, \mnras, 415, L64,
  \dodoi{10.1111/j.1745-3933.2011.01079.x}

\bibitem[{{Martini} {et~al.}(2018){Martini}, {Leroy}, {Mangum}, {Bolatto},
  {Keating}, {Sandstrom}, \& {Walter}}]{martini18}
{Martini}, P., {Leroy}, A.~K., {Mangum}, J.~G., {et~al.} 2018, \apj, 856, 61,
  \dodoi{10.3847/1538-4357/aab08e}

\bibitem[{{McKeith} {et~al.}(1995){McKeith}, {Greve}, {Downes}, \&
  {Prada}}]{mckeith95}
{McKeith}, C.~D., {Greve}, A., {Downes}, D., \& {Prada}, F. 1995, \aap, 293,
  703

\bibitem[{{Melioli} {et~al.}(2013){Melioli}, {de Gouveia Dal Pino}, \&
  {Geraissate}}]{melioli13}
{Melioli}, C., {de Gouveia Dal Pino}, E.~M., \& {Geraissate}, F.~G. 2013,
  \mnras, 430, 3235, \dodoi{10.1093/mnras/stt126}

\bibitem[{{Moran} \& {Lehnert}(1997)}]{moran97}
{Moran}, E.~C., \& {Lehnert}, M.~D. 1997, \apj, 478, 172,
  \dodoi{10.1086/303795}

\bibitem[{{Murray} {et~al.}(2011){Murray}, {M{\'e}nard}, \&
  {Thompson}}]{murray11}
{Murray}, N., {M{\'e}nard}, B., \& {Thompson}, T.~A. 2011, \apj, 735, 66,
  \dodoi{10.1088/0004-637X/735/1/66}

\bibitem[{{Nielsen} {et~al.}(2014){Nielsen}, {Gilfanov}, {Bogd{\'a}n}, {Woods},
  \& {Nelemans}}]{nielsen14}
{Nielsen}, M.~T.~B., {Gilfanov}, M., {Bogd{\'a}n}, {\'A}., {Woods}, T.~E., \&
  {Nelemans}, G. 2014, \mnras, 442, 3400, \dodoi{10.1093/mnras/stu913}

\bibitem[{{Oppenheimer} \& {Dav{\'e}}(2008)}]{oppenheimer08}
{Oppenheimer}, B.~D., \& {Dav{\'e}}, R. 2008, \mnras, 387, 577,
  \dodoi{10.1111/j.1365-2966.2008.13280.x}

\bibitem[{{Oppenheimer} {et~al.}(2016){Oppenheimer}, {Crain}, {Schaye},
  {Rahmati}, {Richings}, {Trayford}, {Tumlinson}, {Bower}, {Schaller}, \&
  {Theuns}}]{oppenheimer16}
{Oppenheimer}, B.~D., {Crain}, R.~A., {Schaye}, J., {et~al.} 2016, \mnras, 460,
  2157, \dodoi{10.1093/mnras/stw1066}

\bibitem[{{Origlia} {et~al.}(2004){Origlia}, {Ranalli}, {Comastri}, \&
  {Maiolino}}]{origlia04}
{Origlia}, L., {Ranalli}, P., {Comastri}, A., \& {Maiolino}, R. 2004, \apj,
  606, 862, \dodoi{10.1086/383018}

\bibitem[{{Peeples} \& {Shankar}(2011)}]{peeples11}
{Peeples}, M.~S., \& {Shankar}, F. 2011, \mnras, 417, 2962,
  \dodoi{10.1111/j.1365-2966.2011.19456.x}

\bibitem[{{Ptak} {et~al.}(1997){Ptak}, {Serlemitsos}, {Yaqoob}, {Mushotzky}, \&
  {Tsuru}}]{ptak97}
{Ptak}, A., {Serlemitsos}, P., {Yaqoob}, T., {Mushotzky}, R., \& {Tsuru}, T.
  1997, \aj, 113, 1286, \dodoi{10.1086/118342}

\bibitem[{{Ranalli} {et~al.}(2008){Ranalli}, {Comastri}, {Origlia}, \&
  {Maiolino}}]{ranalli08}
{Ranalli}, P., {Comastri}, A., {Origlia}, L., \& {Maiolino}, R. 2008, \mnras,
  386, 1464, \dodoi{10.1111/j.1365-2966.2008.13128.x}

\bibitem[{{Read} \& {Stevens}(2002)}]{read02}
{Read}, A.~M., \& {Stevens}, I.~R. 2002, \mnras, 335, L36,
  \dodoi{10.1046/j.1365-8711.2002.05773.x}

\bibitem[{{Rieke} {et~al.}(1980){Rieke}, {Lebofsky}, {Thompson}, {Low}, \&
  {Tokunaga}}]{rieke80}
{Rieke}, G.~H., {Lebofsky}, M.~J., {Thompson}, R.~I., {Low}, F.~J., \&
  {Tokunaga}, A.~T. 1980, \apj, 238, 24, \dodoi{10.1086/157954}

\bibitem[{{Rosen} {et~al.}(2014){Rosen}, {Lopez}, {Krumholz}, \&
  {Ramirez-Ruiz}}]{rosen14}
{Rosen}, A.~L., {Lopez}, L.~A., {Krumholz}, M.~R., \& {Ramirez-Ruiz}, E. 2014,
  \mnras, 442, 2701, \dodoi{10.1093/mnras/stu1037}

\bibitem[{{Roussel} {et~al.}(2010){Roussel}, {Wilson}, {Vigroux}, {Isaak},
  {Sauvage}, {Madden}, {Auld}, {Baes}, {Barlow}, {Bendo}, {Bock}, {Boselli},
  {Bradford}, {Buat}, {Castro-Rodriguez}, {Chanial}, {Charlot}, {Ciesla},
  {Clements}, {Cooray}, {Cormier}, {Cortese}, {Davies}, {Dwek}, {Eales},
  {Elbaz}, {Galametz}, {Galliano}, {Gear}, {Glenn}, {Gomez}, {Griffin}, {Hony},
  {Levenson}, {Lu}, {O'Halloran}, {Okumura}, {Oliver}, {Page}, {Panuzzo},
  {Papageorgiou}, {Parkin}, {Perez-Fournon}, {Pohlen}, {Rangwala}, {Rigby},
  {Rykala}, {Sacchi}, {Schulz}, {Schirm}, {Smith}, {Spinoglio}, {Stevens},
  {Srinivasan}, {Symeonidis}, {Trichas}, {Vaccari}, {Wozniak}, {Wright}, \&
  {Zeilinger}}]{roussel10}
{Roussel}, H., {Wilson}, C.~D., {Vigroux}, L., {et~al.} 2010, \aap, 518, L66,
  \dodoi{10.1051/0004-6361/201014567}

\bibitem[{{Rubin} {et~al.}(2014){Rubin}, {Prochaska}, {Koo}, {Phillips},
  {Martin}, \& {Winstrom}}]{rubin14}
{Rubin}, K. H.~R., {Prochaska}, J.~X., {Koo}, D.~C., {et~al.} 2014, \apj, 794,
  156, \dodoi{10.1088/0004-637X/794/2/156}

\bibitem[{{Salak} {et~al.}(2013){Salak}, {Nakai}, {Miyamoto}, {Yamauchi}, \&
  {Tsuru}}]{salak13}
{Salak}, D., {Nakai}, N., {Miyamoto}, Y., {Yamauchi}, A., \& {Tsuru}, T.~G.
  2013, \pasj, 65, 66, \dodoi{10.1093/pasj/65.3.66}

\bibitem[{{Scannapieco} \& {Br{\"u}ggen}(2015)}]{sca15}
{Scannapieco}, E., \& {Br{\"u}ggen}, M. 2015, \apj, 805, 158,
  \dodoi{10.1088/0004-637X/805/2/158}

\bibitem[{{Schaaf} {et~al.}(1989){Schaaf}, {Pietsch}, {Biermann}, {Kronberg},
  \& {Schmutzler}}]{schaaf89}
{Schaaf}, R., {Pietsch}, W., {Biermann}, P.~L., {Kronberg}, P.~P., \&
  {Schmutzler}, T. 1989, \apj, 336, 722, \dodoi{10.1086/167045}

\bibitem[{{Schneider} {et~al.}(2020){Schneider}, {Ostriker}, {Robertson}, \&
  {Thompson}}]{schneider20}
{Schneider}, E.~E., {Ostriker}, E.~C., {Robertson}, B.~E., \& {Thompson}, T.~A.
  2020, arXiv e-prints, arXiv:2002.10468.
\newblock \doarXiv{2002.10468}

\bibitem[{{Shopbell} \& {Bland-Hawthorn}(1998)}]{shopbell98}
{Shopbell}, P.~L., \& {Bland-Hawthorn}, J. 1998, \apj, 493, 129,
  \dodoi{10.1086/305108}

\bibitem[{{Smith} {et~al.}(2006){Smith}, {Westmoquette}, {Gallagher},
  {O'Connell}, {Rosario}, \& {de Grijs}}]{smith06}
{Smith}, L.~J., {Westmoquette}, M.~S., {Gallagher}, J.~S., {et~al.} 2006,
  \mnras, 370, 513, \dodoi{10.1111/j.1365-2966.2006.10507.x}

\bibitem[{{Smith} {et~al.}(2012){Smith}, {Foster}, \& {Brickhouse}}]{smith12}
{Smith}, R.~K., {Foster}, A.~R., \& {Brickhouse}, N.~S. 2012, Astronomische
  Nachrichten, 333, 301, \dodoi{10.1002/asna.201211673}

\bibitem[{{Socrates} {et~al.}(2008){Socrates}, {Davis}, \&
  {Ramirez-Ruiz}}]{Socrates2008}
{Socrates}, A., {Davis}, S.~W., \& {Ramirez-Ruiz}, E. 2008, \apj, 687, 202,
  \dodoi{10.1086/590046}

\bibitem[{{Strickland} \& {Heckman}(2007)}]{strickland07}
{Strickland}, D.~K., \& {Heckman}, T.~M. 2007, \apj, 658, 258,
  \dodoi{10.1086/511174}

\bibitem[{{Strickland} \& {Heckman}(2009)}]{strickland09}
---. 2009, \apj, 697, 2030, \dodoi{10.1088/0004-637X/697/2/2030}

\bibitem[{{Strickland} {et~al.}(1997){Strickland}, {Ponman}, \&
  {Stevens}}]{strickland97}
{Strickland}, D.~K., {Ponman}, T.~J., \& {Stevens}, I.~R. 1997, \aap, 320, 378.
\newblock \doarXiv{astro-ph/9608064}

\bibitem[{{Strickland} \& {Stevens}(2000)}]{strickland00}
{Strickland}, D.~K., \& {Stevens}, I.~R. 2000, \mnras, 314, 511,
  \dodoi{10.1046/j.1365-8711.2000.03391.x}

\bibitem[{{Suchkov} {et~al.}(1994){Suchkov}, {Balsara}, {Heckman}, \&
  {Leitherer}}]{suchkov94}
{Suchkov}, A.~A., {Balsara}, D.~S., {Heckman}, T.~M., \& {Leitherer}, C. 1994,
  \apj, 430, 511, \dodoi{10.1086/174427}

\bibitem[{{Suchkov} {et~al.}(1996){Suchkov}, {Berman}, {Heckman}, \&
  {Balsara}}]{suchkov96}
{Suchkov}, A.~A., {Berman}, V.~G., {Heckman}, T.~M., \& {Balsara}, D.~S. 1996,
  \apj, 463, 528, \dodoi{10.1086/177267}

\bibitem[{{Tashiro} {et~al.}(2018){Tashiro}, {Maejima}, {Toda}, {Kelley},
  {Reichenthal}, {Lobell}, {Petre}, {Guainazzi}, {Costantini}, {Edison},
  {Fujimoto}, {Grim}, {Hayashida}, {den Herder}, {Ishisaki}, {Paltani},
  {Matsushita}, {Mori}, {Sneiderman}, {Takei}, {Terada}, {Tomida}, {Akamatsu},
  {Angelini}, {Arai}, {Awaki}, {Babyk}, {Bamba}, {Barfknecht}, {Barnstable},
  {Bialas}, {Blagojevic}, {Bonafede}, {Brambora}, {Brenneman}, {Brown},
  {Brown}, {Burns}, {Canavan}, {Carnahan}, {Chiao}, {Comber}, {Corrales}, {de
  Vries}, {Dercksen}, {Diaz-Trigo}, {Dillard}, {DiPirro}, {Done}, {Dotani},
  {Ebisawa}, {Eckart}, {Enoto}, {Ezoe}, {Ferrigno}, {Fukazawa}, {Fujita},
  {Furuzawa}, {Gallo}, {Graham}, {Gu}, {Hagino}, {Hamaguchi}, {Hatsukade},
  {Hawes}, {Hayashi}, {Hegarty}, {Hell}, {Hiraga}, {Hodges-Kluck}, {Holland},
  {Hornschemeier}, {Hoshino}, {Ichinohe}, {Iizuka}, {Ishibashi}, {Ishida},
  {Ishikawa}, {Ishimura}, {James}, {Kallman}, {Kara}, {Katsuda}, {Kenyon},
  {Kilbourne}, {Kimball}, {Kitaguti}, {Kitamoto}, {Kobayashi}, {Kohmura},
  {Koyama}, {Kubota}, {Leutenegger}, {Lockard}, {Loewenstein}, {Maeda},
  {Marbley}, {Markevitch}, {Matsumoto}, {Matsuzaki}, {McCammon}, {McNamara},
  {Miko}, {Miller}, {Miller}, {Minesugi}, {Mitsuishi}, {Mizuno}, {Mori},
  {Mukai}, {Murakami}, {Mushotzky}, {Nakajima}, {Nakamura}, {Nakashima},
  {Nakazawa}, {Natsukari}, {Nigo}, {Nishioka}, {Nobukawa}, {Nobukawa}, {Noda},
  {Odaka}, {Ogawa}, {Ohashi}, {Ohno}, {Ohta}, {Okajima}, {Okamoto}, {Onizuka},
  {Ota}, {Ozaki}, {Plucinsky}, {Porter}, {Pottschmidt}, {Sato}, {Sato},
  {Sawada}, {Seta}, {Shelton}, {Shibano}, {Shida}, {Shidatsu}, {Shirron},
  {Simionescu}, {Smith}, {Someya}, {Soong}, {Suagawara}, {Szymkowiak},
  {Takahashi}, {Tamagawa}, {Tamura}, {Tanaka}, {Terashima}, {Tsuboi},
  {Tsujimoto}, {Tsunemi}, {Tsuru}, {Uchida}, {Uchiyama}, {Ueda}, {Uno},
  {Walsh}, {Watanabe}, {Williams}, {Wolfs}, {Wright}, {Yamada}, {Yamaguchi},
  {Yamaoka}, {Yamasaki}, {Yamauchi}, {Yamauchi}, {Yanagase}, {Yaqoob},
  {Yasuda}, {Yoshioka}, {Zabala}, \& {Irina}}]{tashiro18}
{Tashiro}, M., {Maejima}, H., {Toda}, K., {et~al.} 2018, in Society of
  Photo-Optical Instrumentation Engineers (SPIE) Conference Series, Vol. 10699,
  \procspie, 1069922

\bibitem[{{The Lynx Team}(2018)}]{lynx18}
{The Lynx Team}. 2018, arXiv e-prints, arXiv:1809.09642.
\newblock \doarXiv{1809.09642}

\bibitem[{{Thompson} {et~al.}(2015){Thompson}, {Fabian}, {Quataert}, \&
  {Murray}}]{thompson15}
{Thompson}, T.~A., {Fabian}, A.~C., {Quataert}, E., \& {Murray}, N. 2015,
  \mnras, 449, 147, \dodoi{10.1093/mnras/stv246}

\bibitem[{{Thompson} {et~al.}(2016){Thompson}, {Quataert}, {Zhang}, \&
  {Weinberg}}]{thompson16}
{Thompson}, T.~A., {Quataert}, E., {Zhang}, D., \& {Weinberg}, D.~H. 2016,
  \mnras, 455, 1830, \dodoi{10.1093/mnras/stv2428}

\bibitem[{{Tsuru} {et~al.}(1997){Tsuru}, {Awaki}, {Koyama}, \&
  {Ptak}}]{tsuru97}
{Tsuru}, T.~G., {Awaki}, H., {Koyama}, K., \& {Ptak}, A. 1997, \pasj, 49, 619,
  \dodoi{10.1093/pasj/49.6.619}

\bibitem[{{Tsuru} {et~al.}(2007){Tsuru}, {Ozawa}, {Hyodo}, {Matsumoto},
  {Koyama}, {Awaki}, {Fujimoto}, {Griffiths}, {Kilbourne}, {Matsushita},
  {Mitsuda}, {Ptak}, {Ranalli}, \& {Yamasaki}}]{tsuru07}
{Tsuru}, T.~G., {Ozawa}, M., {Hyodo}, Y., {et~al.} 2007, \pasj, 59, 269,
  \dodoi{10.1093/pasj/59.sp1.S269}

\bibitem[{{Umeda} {et~al.}(2002){Umeda}, {Nomoto}, {Tsuru}, \&
  {Matsumoto}}]{umeda02}
{Umeda}, H., {Nomoto}, K., {Tsuru}, T.~G., \& {Matsumoto}, H. 2002, \apj, 578,
  855, \dodoi{10.1086/342650}

\bibitem[{{Veilleux} {et~al.}(2005){Veilleux}, {Cecil}, \&
  {Bland-Hawthorn}}]{veilleux05}
{Veilleux}, S., {Cecil}, G., \& {Bland-Hawthorn}, J. 2005, \araa, 43, 769,
  \dodoi{10.1146/annurev.astro.43.072103.150610}

\bibitem[{{Verner} {et~al.}(1996){Verner}, {Ferland}, {Korista}, \&
  {Yakovlev}}]{verner96}
{Verner}, D.~A., {Ferland}, G.~J., {Korista}, K.~T., \& {Yakovlev}, D.~G. 1996,
  \apj, 465, 487, \dodoi{10.1086/177435}

\bibitem[{{Walter} {et~al.}(2002){Walter}, {Weiss}, \& {Scoville}}]{walter02}
{Walter}, F., {Weiss}, A., \& {Scoville}, N. 2002, \apjl, 580, L21,
  \dodoi{10.1086/345287}

\bibitem[{{Watson} {et~al.}(1984){Watson}, {Stanger}, \&
  {Griffiths}}]{watson84}
{Watson}, M.~G., {Stanger}, V., \& {Griffiths}, R.~E. 1984, \apj, 286, 144,
  \dodoi{10.1086/162583}

\bibitem[{{Werk} {et~al.}(2016){Werk}, {Prochaska}, {Cantalupo}, {Fox},
  {Oppenheimer}, {Tumlinson}, {Tripp}, {Lehner}, \& {McQuinn}}]{werk16}
{Werk}, J.~K., {Prochaska}, J.~X., {Cantalupo}, S., {et~al.} 2016, \apj, 833,
  54, \dodoi{10.3847/1538-4357/833/1/54}

\bibitem[{{Westmoquette} {et~al.}(2009){Westmoquette}, {Gallagher}, {Smith},
  {Trancho}, {Bastian}, \& {Konstantopoulos}}]{westmoquette09}
{Westmoquette}, M.~S., {Gallagher}, J.~S., {Smith}, L.~J., {et~al.} 2009, \apj,
  706, 1571, \dodoi{10.1088/0004-637X/706/2/1571}

\bibitem[{{Westmoquette} {et~al.}(2007){Westmoquette}, {Smith}, {Gallagher},
  {O'Connell}, {Rosario}, \& {de Grijs}}]{westmoquette07}
{Westmoquette}, M.~S., {Smith}, L.~J., {Gallagher}, J.~S., I., {et~al.} 2007,
  \apj, 671, 358, \dodoi{10.1086/522693}

\bibitem[{{Zhang} {et~al.}(2017){Zhang}, {Thompson}, {Quataert}, \&
  {Murray}}]{zhang17}
{Zhang}, D., {Thompson}, T.~A., {Quataert}, E., \& {Murray}, N. 2017, \mnras,
  468, 4801, \dodoi{10.1093/mnras/stx822}

\bibitem[{{Zhang} {et~al.}(2014){Zhang}, {Wang}, {Ji}, {Smith}, {Foster}, \&
  {Zhou}}]{zhang14}
{Zhang}, S., {Wang}, Q.~D., {Ji}, L., {et~al.} 2014, \apj, 794, 61,
  \dodoi{10.1088/0004-637X/794/1/61}

\end{thebibliography}

\end{document}